\documentclass[prd,eqsecnum,amsfonts,showpacs,draft,twocolumn,superscriptaddress,
hyperref,nofootinbib]{revtex4}

\usepackage{graphicx}
\usepackage{amsfonts}
\usepackage{bm}
\usepackage{hyperref}

\usepackage{color}
\definecolor{IITred}{rgb}{0.5,0.05,0.05}

\newcommand{\SMR}{$\overline{\mathrm{SM}}$}
\newcommand{\SMRn}[1]{$\overline{\mathrm{SM}}_{#1}$}
\newcommand{\SMRm}{$\overline{\mathrm{SM}}m$}
\newcommand{\LR}{$\overline{\mathrm{LR}}$}
\newcommand{\PS}{$\overline{\mathrm{PS}}$}

\def\bra#1{\langle #1|}
\def\ket#1{| #1\rangle}
\newcommand{\uup}{\ensuremath{u_\uparrow}}
\newcommand{\dup}{\ensuremath{d_\uparrow}}
\newcommand{\udn}{\ensuremath{u_\downarrow}}
\newcommand{\ddn}{\ensuremath{d_\downarrow}}

\newcommand{\ie}{{\em i.e.}}
\newcommand{\cf}{{\em cf.\ }}

\newcommand{\gev}{\hbox{ GeV}}

\newcommand{\mev}{\hbox{ MeV}}

\newcommand{\tev}{\hbox{ TeV}}

\newcommand{\s}{\hbox{ s}}
\newcommand{\ps}{\hbox{ ps}}
\newcommand{\fm}{\hbox{ fm}}
\newcommand{\cm}{\hbox{ cm}}

\newcommand{\eqn}[1]{(\ref{#1})}
\newcommand{\Eqn}[1]{Eq.~(\ref{#1})}
\newcommand{\abs}[1]{\left| #1\right|}
\newcommand{\smgg}{\ensuremath{\mathrm{SU(3)_c}\otimes \mathrm{SU(2)_L}\otimes \mathrm{U(1)}_Y}}
\newcommand{\lrgg}{\ensuremath{\mathrm{SU(3)_c}\otimes \mathrm{SU(2)_L}\otimes \mathrm{SU(2)_R}\otimes \mathrm{U(1)}_{B-L}}}
\newcommand{\psew}{\ensuremath{\mathrm{SU(4)_{PS}}\otimes \mathrm{SU(2)_L}\otimes \mathrm{SU(2)_R}}}
\newcommand{\psgg}{\ensuremath{\mathrm{SU(4)_{PS}}}}
\newcommand{\ewgg}{\ensuremath{\mathrm{SU(2)_L}\otimes \mathrm{U(1)}_Y}}
\newcommand{\cgg}{\ensuremath{\mathrm{SU(3)_c}}}
\newcommand{\wigg}{\ensuremath{\mathrm{SU(2)_L}}}
\newcommand{\emgg}{\ensuremath{\mathrm{U(1)}_{\mathrm{em}}}}
\newcommand{\blgg}{\ensuremath{\mathrm{U(1)}_{B-L}}}
\newcommand{\chiral}[1]{\ensuremath{\mathrm{SU}(#1)_{\mathrm{L}} \otimes \mathrm{SU}(#1)_{\mathrm{R}}}}

\def\vev#1{\left\langle #1\right\rangle_0}
\newcommand{\cfrac}[2]{\ensuremath{\textstyle{\frac{#1}{#2}}}}

\newcommand{\beq}{\begin{equation}}
\newcommand{\eeq}{\end{equation}}
\newcommand{\beqs}{\begin{eqnarray}}
\newcommand{\eeqs}{\end{eqnarray}}

\newcommand{\fund}{\ensuremath{\mathbf{n_q}}}
        
\preprint{FERMILAB--PUB--09/018--T}
\preprint{YITP-SB-08-32}

\begin{document}

\title{\textit{Gedanken} Worlds without Higgs: \\
QCD-Induced Electroweak Symmetry Breaking}

\author{Chris Quigg}

\affiliation{Theoretical Physics Department \\ Fermi National Accelerator Laboratory, 
 Batavia, Illinois 60510 USA}
\affiliation{Institut f\"{u}r Theoretische Teilchenphysik \\ Universit\"{a}t Karlsruhe, D-76128 Karlsruhe, Germany}

\author{Robert Shrock}

\affiliation{C.~N.~Yang Institute for Theoretical Physics \\
Stony Brook University, 
Stony Brook, New York 11794 USA}

\begin{abstract}
To illuminate how electroweak symmetry breaking shapes the physical world, we investigate toy models in which no Higgs fields or other constructs are introduced to induce spontaneous symmetry breaking.
 Two models incorporate the standard \smgg\ gauge symmetry and fermion content similar to that of the standard model. The first class---like the standard electroweak theory---contains no bare mass terms, so the spontaneous breaking of chiral symmetry within quantum chromodynamics is the only source of electroweak symmetry breaking. The second class adds bare fermion masses sufficiently small that QCD remains the dominant source of electroweak symmetry breaking and  the model can serve as a well-behaved low-energy effective field theory to energies somewhat above the hadronic scale. 
 A third class of models is based on the left-right--symmetric \lrgg\ gauge group. In a fourth class of models, built on \psew\ gauge symmetry, lepton number is treated as a fourth color.
 Many interesting characteristics of the models stem from the fact that  the effective strength of the weak interactions is much closer to that of the residual strong interactions than in the real world. 
 The Higgs-free models not only provide informative contrasts to the real world, but also lead us to consider intriguing issues in the application of field theory to the real world.
 
\end{abstract}
\pacs{11.15.-q, 12.10-g, 12.60.-i \hfill\textsf{YITP-SB-08-32 \hfill FERMILAB--PUB--09/018--T}}

\maketitle

\section{Introduction} 
Over the past fifteen years, the electroweak theory~\cite{Weinberg:1967tq} has been elevated from a promising description to a provisional law of nature, tested as a quantum field theory at the level of one per mille by many measurements~\cite{EWWG}. Joined with quantum chromodynamics, the theory of the strong interactions, to form the \textit{standard model} based on the gauge group \smgg, and augmented to incorporate neutrino masses and lepton mixing, it describes a vast array of experimental information.

In this picture, the electroweak symmetry is spontaneously broken, $\ewgg \to \mathrm{U(1)_{em}}$, when an elementary complex scalar field $\phi$ that transforms as a (color-singlet) weak-isospin doublet with weak hypercharge $Y_\phi = 1$ acquires a nonzero vacuum expectation value, by virtue of its self-interactions~\cite{Englert:1964et}. The scalar field is introduced as the agent of electroweak symmetry breaking and its self-interactions, given by the potential
$V(\phi^\dagger \phi) = \mu^2(\phi^\dagger \phi) +
\abs{\lambda}(\phi^\dagger \phi)^2$,
are arranged so that the vacuum state corresponds to a 
broken-symmetry solution.  The electroweak symmetry is spontaneously broken if the parameter
$\mu^2$ is taken to be negative. In that event, gauge invariance gives us the freedom to choose the state of minimum energy---the vacuum state---to correspond to the vacuum expectation value
\begin{equation}
\vev{\phi} = \vev{\left(\begin{array}{c} \phi^+ \\ \phi^0 \end{array} \right)} = 
\left(\begin{array}{c}0 \\ v/\sqrt{2} \end{array} \right),
\end{equation}
where $v = \sqrt{-\mu^2/\!\abs{\lambda}}$. Three of the four degrees of freedom of $\phi$ and $\phi^\dagger$ become the longitudinal components of the gauge bosons $W^+, W^-, Z^0$. The fourth emerges as a massive scalar particle $H$, called the Higgs boson, with its mass given symbolically by $M_H^2 = -2\mu^2 = \sqrt{2\!\abs{\lambda}}v$.

Fits to a universe of electroweak precision measurements~\cite{EWWG} are in excellent agreement with the standard model. However, the Higgs boson has not been observed directly, and we do not know whether such a fundamental field exists or whether some different mechanism breaks electroweak symmetry. One of the great campaigns now under way in both experimental and theoretical particle physics is to advance our understanding of electroweak symmetry breaking by finding $H$ or its stand-in.

For all its successes, the electroweak theory leaves many questions unanswered. It does not explain the choice $\mu^2 < 0$ required to hide the electroweak symmetry, and it merely accommodates, but does not predict, fermion masses and mixings. Moreover, the Higgs sector is unstable against large radiative corrections. A second great campaign has been to imagine more complete and predictive extensions to the electroweak theory, and to test for experimental signatures of those extensions, which include supersymmetry, dynamical symmetry breaking, and the influence of extra spacetime dimensions. These more ambitious theories also put forward tentative answers to questions that lie beyond the scope of the standard model: the nature of dark matter, the matter asymmetry of the universe, etc. {Theories that incorporate quarks and leptons into extended families point toward unification of the separate \smgg\ gauge couplings. They may also provide a rationale for charge quantization, open paths to relationships among the masses of standard-model fermions, and even determine the number of fermion generations.}

The aim of this study is more modest, and more pedagogical. We step back from the task of constructing a realistic model of particle physics valid to the TeV scale and beyond, and address instead why electroweak symmetry breaking matters for the physical world. Our approach is to analyze closely what the world would be like in the absence of electroweak symmetry breaking at the usual scale ($v = 2^{-1/4}\,G_{\mathrm{F}}^{-{1}/{2}} \approx 246\gev$), whether by the conventional Higgs mechanism or by any of its alternatives, including dynamical symmetry breaking and higher-dimensional formulations. Our laboratory will consist of four classes of toy models. Two are drawn from the standard model by eliminating the Higgs sector, a third is derived from a left-right--symmetric electroweak gauge symmetry, and the fourth joins a left-right--symmetric electroweak symmetry to the Pati--Salam~\cite{Pati:1974yy} symmetry \psgg, in which lepton number is identified as a fourth color.

Note that the program reported here is quite distinct from the attempt to construct ``Higgsless'' models of the real world~\cite{Hill:2000mu}, by introducing alternative mechanisms for electroweak symmetry breaking. It is unrelated to attempts to quantify the extent to which parameters of the standard model might be environmentally selected~\cite{Hogan:1999wh}. Our interest is to determine how certain \textit{Gedanken} worlds that follow well-defined rules differ from the real world. Our approach therefore shares something of the inquisitive spirit of varying the vacuum expectation value $v$ while holding fixed other parameters, such as the Yukawa couplings that set fermion masses~\cite{Agrawal:1997gf},  or considering \textit{what if?} variations of standard-model parameters~\cite{Cahn:1996ag}.\footnote{This broad area of investigation has an abundant literature; see~\cite{anth} for a sampler. Some early related discussions that focus on cosmological issues but also consider the variation of particle physics parameters are noted in~\cite{cosmoanth}.} However, we use standard field-theoretic calculational methods, rather than anthropic arguments, to reach our conclusions.

We designate the simplest class of models that we investigate as the modified standard model, \SMR. Models in this class are built on the standard-model gauge group, \smgg, with gauge couplings similar to the standard ones, and similar fermion content. The essential difference from the standard-model setup is the absence of a Higgs sector, or any other element contrived to hide the electroweak symmetry at the usual scale of $246\gev$.

Even a one-generation modified standard model provides a highly useful context for illustrating what an understanding of the agent of electroweak symmetry breaking will reveal about the everyday world~\cite{Quigg:2007dt}. It is worth previewing a few features here to indicate how different that \SMR\ \textit{Gedanken} world would be. Eliminating the Higgs mechanism does nothing to alter the strong interaction, so QCD would
still confine color-triplet quarks and color-octet gluons into color-singlet hadrons.  

If the electroweak symmetry remained unbroken, the asymptotically free $\mathrm{SU(2)_L}$ force would, in similar fashion, confine objects that carry weak isospin into weak-isospin singlets. But as we shall recall in \S\ref{sec:SMR}, QCD spontaneously breaks the chiral symmetry of the massless $u$ and $d$ quarks, forming a quark condensate that hides the electroweak symmetry, even in the absence of a Higgs mechanism. 
Because the weak bosons have acquired mass by absorbing the would-be pions, the $\mathrm{SU(2)_L}$ interaction does not confine. Quarks and leptons would
remain massless, because QCD-induced electroweak symmetry breaking has no analogue of the Higgs-boson Yukawa terms of the standard model. Without a Higgs boson, the interactions among weak gauge bosons become strongly coupled.

Apart from the absent pions, the familiar light-hadron spectrum persists, but with crucial differences. 
In the \SMR, no quark masses means no $u$-$d$ quark mass difference to overcome the larger electromagnetic self-energy of the proton and make the neutron outweigh the proton. Instead, the outcome depends on a competition between $\gamma$- and $Z$-induced mass shifts that leads to $\abs{M_n - M_p} \approx \hbox{a few}\mev$. A possible outcome is that the pattern of radioactive beta decay might be turned on its head. In the real world, a free neutron decays, $n \to p e^- \bar{\nu}_e$, with a mean life of $885.7 \pm 0.8\s$, about fifteen minutes. If the $n$-$p$ mass difference changed sign and the $W$-boson mass were characteristic of the QCD scale, a free proton would decay in about 15 picoseconds: $p \to n e^+ \nu_e$. There would be no hydrogen atom, and the lightest ``nucleus'' would be one neutron. 

Straightforward modifications of big-bang nucleosynthesis hint that some light elements would be produced in the early no-Higgs universe~\cite{Hogan:1999wh,Agrawal:1997gf,Yoo:2002vw}. But even if some nuclei were produced and survived, they would not form recognizable atoms. A massless electron means that the Bohr radius of an atom---half a nanometer in the real world---would be infinite.\footnote{Now, it is nearly inevitable that effects negligible in our world would, in the Higgsless world, produce fermion masses many orders of magnitude smaller than those we observe. The Bohr radius of a would-be atom would be macroscopic, sustaining the conclusion that matter would lose its integrity.} In a world without compact atoms, valence chemical bonding would have no meaning.  All matter would be insubstantial---and life as we know it would not exist!  On top of all that, the vacuum would be
unstable to the formation of a plasma of $e^+e^-$ pairs.  
The minimalist \textit{Gedanken} world gives eloquent testimony to the importance of electroweak symmetry breaking for the properties of our world.

More can be learned, as we shall see, from a close analysis of the \SMR, but the vacuum instability for massless electrons is a sign of pathological behavior. Moreover, we wish to regard the \SMR\ as a low-energy effective field theory that could be obtained by integrating out massive degrees of freedom from a more ultraviolet-complete theory. It is natural then to imagine that generic higher-dimension operators would induce bare fermion masses that are hard on the QCD scale set by $\Lambda_{\mathrm{QCD}}$. Accordingly, we consider a second class of \textit{Gedanken} worlds, the modified standard model with bare fermion masses, abbreviated \SMRm. Mass terms added by fiat must, of course, respect the surviving $\mathrm{SU(3)_c} \otimes \mathrm{U(1)_{em}}$ gauge symmetry of the low-energy standard model. 

Specific mechanisms analogous to extended technicolor can give rise to bare mass terms of the kind we wish to entertain, but we shall not specify the elements in the ultraviolet completion that produce the fermion masses. The bare fermion mass terms break the electroweak symmetry explicitly. The requirement that such models serve as consistent low-energy effective field theories up to a specified energy places an upper bound on the admissible bare fermion masses.
We note that judiciously combining small explicit symmetry breaking with spontaneous symmetry breaking has been rewarded with insights in the past, among them the realization that pions can be understood as (what we now call) pseudo-Nambu--Goldstone bosons~\cite{Nambu:1960xd}.

 The third class of models we examine is based on the left-right--symmetric \lrgg\ gauge group, where $B$ and $L$ denote baryon number and lepton number, respectively. In the real-world context, this gauge group is motivated by the hypothesis that the weak interactions are, at bottom, left-right symmetric, and that the observed parity violation is a consequence of the pattern of electroweak symmetry breaking. As we shall review in \S\ref{sec:lrsym}, the electric charge operator has a more symmetric form than in the standard \ewgg\ electroweak theory, and the seeming arbitrariness of fermion weak-hypercharge assignments is removed. In the left-right--symmetric \textit{Gedanken} world, both leptons and quarks are confined, so there is no pathology associated with vanishing lepton masses. 
 
 The final example we study is a prototype unified theory, in which the \cgg\ and $\mathrm{U(1)}_{B-L}$ symmetries are combined into an \psgg\  ``Pati-Salam'' symmetry. The \psew\ model explains the quantization of electric charge. We will show in \S\ref{sec:psworld} that in the Pati-Salam \textit{Gedanken} world, all the fermions pick up dynamical masses at the \psgg\ confinement scale. There is no infrared pathology provoked by the existence of massless deconfined particles. The low-energy effective field theory exhibits a residual exact $\mathrm{SU(2)_V}$ gauge symmetry that confines, producing a spectrum of light counterparts of QCD glueballs. No exact Abelian symmetry emerges, so the low-energy effective theory contains no analogue of electromagnetism.
 
We explore the \SMR\ class of models in \S\ref{sec:SMR}.
We shall devote most of our analysis to the simplest modified standard model with a single generation of massless fermions. We look first at the electroweak symmetry breaking mediated by the strong interactions. Then we consider the hadron spectrum in the \textit{Gedanken} world and examine changes to the real-world pattern of $\beta$ decay. In a model without a Higgs boson, scattering of weak gauge bosons becomes strong at modest c.m.\ energies, below $1\gev$. We remark on changes to strong-interaction symmetries in the  \textit{Gedanken} world, and on the form of the long-range nuclear interactions, which is very different from the one-pion-exchange potential of the real world.

Then we extend the model to include multiple generations of massless fermions. We derive scaling relations for electroweak observables and remark on qualitative changes to the hadron spectrum that accompany the introduction of quarks beyond the first generation. 

To circumvent the infrared pathologies brought about by massless charged fermions, we introduce explicit quark and lepton masses in \S\ref{sec:baremass}. We derive conditions on the magnitude of those masses, and note changes with respect to the models with massless fermions. In \S\ref{sec:colors} we look briefly at generalizations to the number of quark colors, particularly in the limit of large $N_{\mathrm{c}}$. Section~\ref{sec:lrsym} explores the implications of a left-right symmetric gauge group for a \textit{Gedanken} world without the Higgs mechanism. The extension to an example left-right--symmetric unifying group occupies \S\ref{sec:psworld}. 
A concluding \S\ref{sec:conc} summarizes what we have learned.

\section{No Higgs Mechanism, \newline No Bare Fermion Masses \label{sec:SMR}}
Lessons from the superconducting phase transition informed the understanding of spontaneous symmetry breaking on the subatomic scale~\cite{Nambu:1960tm,Goldstone:1961eq} and led to the insight that the spontaneous breaking of a local gauge symmetry gives masses to gauge bosons~\cite{Englert:1964et}. The spontaneous symmetry breaking in the standard electroweak theory~\cite{Weinberg:1967tq} is modeled on the Ginzburg--Landau description of superconductivity~\cite{Ginzburg:1950sr}.\footnote{See Ref.~\cite{Kibble:2009} for a brief, authoritative account of the path from symmetry breaking in superconductivity to massive gauge bosons in relativistic quantum field theory.} In the Ginzburg--Landau phenomenology, the macroscopic order parameter, which corresponds to the wave function of superconducting charge carriers, acquires a nonzero vacuum expectation value in the 
superconducting state. Within a superconductor, the photon picks up a mass $M_\gamma = \hbar/\lambda_{\mathrm{L}}$, where $\lambda_{\mathrm{L}}$ is the London penetration depth that characterizes the exclusion of magnetic flux by the Meissner effect. In like manner, the auxiliary scalars introduced to hide the electroweak symmetry acquire a nonzero vacuum expectation value. The weak gauge bosons become massive in the process.

In the case of superconductivity, we have an example of a gauge-symmetry--breaking mechanism that does not rely on introducing an \textit{ad hoc} order parameter.
The microscopic Bardeen-Cooper-Schrieffer 
theory \cite{Bardeen:1957mv} reveals a dynamical origin of the order parameter through 
the formation of correlated states of elementary fermions, the Cooper pairs of 
electrons. This suggests that the electroweak symmetry might also be broken dynamically, without the need to introduce scalar fields. Indeed, quantum chromodynamics can be the source of electroweak symmetry breaking.

\subsection{One fermion generation \label{subsec:onegen}}

Let us consider first, as a low-energy effective field theory, a modified standard model \SMRn{1} with a single generation of massless fermions and no scalar sector: an \smgg\ theory of massless up and 
down quarks, plus massless electron and neutrino. The chiral quark fields are
\begin{equation}
\mathsf{Q}^a_{\mathrm{L}} = \left(\begin{array}{c} u^{a} \\ d^a\end{array}\right)_{\!\mathrm{L}}, \qquad u^a_{\mathrm{R}}, \quad d^a_{\mathrm{R}}\;,
\label{eq:chiralq}
\end{equation}
with $(\cgg,\wigg)_Y$ quantum numbers $(\mathbf{3},\mathbf{2})_{1/3}$, $(\mathbf{3},\mathbf{1})_{4/3}$, and $(\mathbf{3},\mathbf{1})_{-2/3}$, respectively, and color index $a = 1,2,3$. The chiral lepton fields are 
\begin{equation}
\mathsf{L}_{\mathrm{L}} = \left(\begin{array}{c} \nu \\ e\end{array}\right)_{\!\mathrm{L}}, \qquad e_{\mathrm{R}}\;,
\label{eq:chirall}
\end{equation}
with quantum numbers $(\mathbf{1},\mathbf{2})_{-1}$ and $(\mathbf{1},\mathbf{1})_{-2}$, respectively. Electroweak-singlet ``sterile'' neutrinos $N_{\mathrm{R}}$, with $(\mathbf{1},\mathbf{1})_{0}$ quantum numbers may be added at will.

Provided that the \cgg\ color gauge interaction is dominant, so that we may treat the \ewgg\
interactions as a perturbation, these models have
the striking property that the QCD quark condensates dynamically break
electroweak symmetry.  
\subsubsection{Electroweak symmetry breaking}
For vanishing quark masses (and with electroweak interactions turned off), QCD displays an exact 
$\mathrm{SU(2)_L\otimes SU(2)_R}$ chiral symmetry. At an energy scale 
$\sim\Lambda_{\mathrm{QCD}},$ the strong interactions become strong and $\langle \bar{q}q \rangle \equiv \langle\sum_{a=1}^{N_{\mathrm{c}}}\bar{q}^a q^a\rangle$ quark 
condensates appear. Vacuum-alignment arguments~\cite{Dashen:1970et} imply that the condensates are not only color-neutral, but also electrically neutral, of the form $\langle\bar{u}u\rangle$ and $\langle\bar{d}d\rangle$. In the limit of vanishing \ewgg\ interactions, $\langle\bar{u}u\rangle=\langle\bar{d}d\rangle$. Formation of the quark condensates spontaneously breaks the chiral symmetry to the familiar isospin flavor symmetry,\footnote{The chiral symmetry breaking is associated with the dynamical generation of equal ``constituent'' masses for the up and down quarks.} 
\begin{equation}
	\mathrm{SU(2)_L\otimes SU(2)_R \to SU(2)_V}\;\; .
\end{equation}
 Three Nambu--Goldstone bosons appear, one for 
each broken generator of the original chiral invariance. These were 
identified by Nambu~\cite{Nambu:1960xd} as three massless pions.

The $\langle\bar q q\rangle = \langle\bar q_{\mathrm{L}} q_{\mathrm{R}} + \bar q_{\mathrm{R}} q_{\mathrm{L}}\rangle$ condensate links left-handed and right-handed quarks, which transform differently under \ewgg. It follows from the weak-isospin and weak-hypercharge quantum numbers of the left- and right-handed $u$ and $d$ quarks that the condensate transforms as a weak isodoublet with weak hypercharge $|Y|=1$,
and hence breaks $\ewgg \to \emgg$.

The broken generators correspond to three axial currents whose couplings to pions are 
measured by the pion decay constant $f_\pi$, which is defined through the matrix element
\begin{equation}
\bra{0}J_j^\nu\ket{\pi_k(q)} = i \delta_{jk}f_\pi q^\nu,
\label{eq:fpidef}
\end{equation}
by which the charged weak current $J_j^\nu$ connects the pion to the vacuum. 
The measured value of the charged-pion lifetime determines the real-world value of the decay constant, $f_{\pi} \approx 92.4\mev$~\cite{Amsler:2008zz}. Within chiral perturbation theory, Gasser and Leutwyler~\cite{Gasser:1984gg} have estimated that $f_\pi$ would decrease by about 6\% relative to its experimental value  if the $u$ and $d$ quark masses were reduced to zero from the real-world value, $\case{1}{2}(m_u + m_d) = 2.5\hbox{ to }5\mev$~\cite{Amsler:2008zz}. We adopt $\bar{f}_\pi \approx  87\mev$ as an estimate of the corresponding parameter in the \SMRn{1}.\footnote{A superior bar denotes quantities in the \SMR\ \textit{Gedanken} world.}  

When we turn on the 
\ewgg\ electroweak interaction with couplings $g$ and $g^\prime$, the electroweak gauge 
bosons couple to the axial currents and acquire masses of order $\sim 
g\bar{f}_\pi$. In the interplay between the electroweak gauge interactions and QCD, the would-be massless pions disappear from the hadron spectrum, 
having become the longitudinal components of the weak gauge bosons. 
The mass-squared matrix,
\begin{equation}
	\mathcal{M}^{2} = \left(
		\begin{array}{cccc}
		g^{2} & 0 & 0 & 0  \\
		0 & g^{2} & 0 & 0  \\
		0 & 0 & g^{2} & gg^{\prime}  \\
		0 & 0 & gg^{\prime} & g^{\prime2}
	\end{array}
		 \right) \frac{\bar{f}_{\pi}^{2}}{4} \; ,
	\label{eq:csbm2}
\end{equation}
(where the rows and columns correspond to the three weak-isospin gauge bosons $b_1$, $b_2$, $b_{3}$, and the weak-hypercharge gauge boson $\mathcal{A}$) has the same structure as the mass-squared matrix 
for gauge bosons in the standard electroweak theory.  

Diagonalizing 
the matrix \eqn{eq:csbm2}, we find that the photon, corresponding as in the standard model to the combination $A = (g\mathcal{A} + g^{\prime}b_3)/\sqrt{g^2 + g^{\prime 2}}$, emerges massless.
Two charged gauge bosons, $W^{\pm} = (b_1 \mp i b_2)/\sqrt{2}$, acquire mass-squared
$\overline{M}_{W}^{2} = g^{2}\bar{f}_{\pi}^{2}/4$, and the neutral gauge boson, $Z = (-g^{\prime}\mathcal{A} + gb_3)/ \sqrt{g^2 + g^{\prime 2}}$, obtains $\overline{M}_{Z}^{2} = 
(g^{2}+g^{\prime2})\bar{f}_{\pi}^{2}/4$.The ratio,
\begin{equation}
	{\overline{M}_{Z}^{2}}/{\overline{M}_{W}^{2}} ={(g^{2}+g^{\prime2})}/{g^{2}} = 
	{1}/{\cos^{2}\theta_{\mathrm{W}}}\; ,
	\label{eq:wzrat}
\end{equation}
where $\theta_{\mathrm{W}}$ is the weak mixing angle, echos the standard-model result, because the \ewgg\ transformation properties of the quark condensate ensure a custodial $\mathrm{SU(2)}$ symmetry.

Here the symmetry breaking is dynamical and automatic; it can be traced, through spontaneous chiral symmetry breaking and confinement, to the asymptotic freedom of QCD. Electroweak symmetry breaking determined by pre-existing dynamics stands in
contrast to the standard electroweak theory, in which spontaneous symmetry breaking results from the
{\it ad hoc} choice of $\mu^2 < 0$  for the coefficient of the quadratic
term in the Higgs potential.

Despite the structural similarity to the standard model, the chiral symmetry breaking of QCD does not yield a satisfactory theory of the weak interactions. To illustrate this statement quantitatively, we must specify the values of the gauge couplings. In the case of QCD, we have already assumed that the scale $\Lambda_{\mathrm{QCD}}$ retains its real-world value, which corresponds to $\alpha_3(M_Z) \approx 0.118$, i.e., $g_3(M_Z) \approx 1.22$. In that spirit, we shall take the values of the \ewgg\ couplings at $\overline{M}_{Z}$ to be the same as those measured in the real world at $M_Z$: $g \approx 0.65$ and $g^\prime \approx 0.34$~\cite{Amsler:2008zz}.\footnote{From the perspective of a unified theory, the low-energy values of the gauge couplings are determined by renormalization-group evolution down from the unification scale. We do not specify an ultraviolet completion for the \SMR, so the low-energy values of $g$ and $g^\prime$ are not specified unambiguously. Our qualitative conclusions do not depend sensitively on the precise values we adopt.}

The masses acquired by the 
intermediate bosons are some $2\,800$ times smaller than required for a successful 
low-energy phenomenology, because their scale is set by $\bar{f}_{\pi} \approx 87\mev$, instead of $v \approx 246\gev$~\cite{Weinstein:1973gj}:
\begin{equation}
\overline{M}_W \approx 28\mev; \quad\overline{M}_Z \approx 32\mev.
\label{eq:SMRWmass}
\end{equation}
The change in scale from the real-world values of $M_W = 80.398 \pm 0.025\gev$ and $M_Z = 91.1876 \pm 0.0021\gev$~\cite{Amsler:2008zz} has many far-reaching implications.

The modified standard model \SMRn{1}\  serves as a reminder that
 hiding the electroweak symmetry and generating masses for $W^\pm$ and
$Z^0$ masses does not, by itself, confer masses on the fermions.\footnote{To be sure, quarks do gain the usual dynamical (``constituent'') masses from their color interactions and associated confinement. The constituent masses are soft, i.e., the associated running masses decay to zero at Euclidean momenta far above the scale of confinement and spontaneous chiral symmetry
breaking, set by $\Lambda_{\mathrm{QCD}} \simeq 200$ MeV. However, the quarks do not acquire
any hard masses, i.e., masses that persist for momenta large compared with
$\Lambda_{\mathrm{QCD}}$. The charged leptons and neutrinos remain massless in every sense, to this approximation.} The possible division of labor between electroweak symmetry breaking and the generation of gauge-boson masses on the one hand, and the generation of fermion masses on the other hand is to be kept in mind as we explore the TeV scale.

The observation that QCD dynamically breaks electroweak symmetry (but at too low a scale) inspired the invention of analogous no-Higgs theories in which dynamical symmetry breaking is accomplished by  the formation of a condensate of new fermions subject to a new, asymptotically free, vectorial gauge interaction (often called
technicolor) that becomes strongly coupled at the TeV scale~\cite{Weinberg:1979bn}. 
 The technifermion condensates that
dynamically break the electroweak symmetry produce masses for the $W^\pm$ and $Z^0$
bosons but do not directly give masses to the standard-model fermions.  To endow the quarks and leptons with mass, it is necessary to embed technicolor in a larger \textit{extended technicolor}
framework containing degrees of freedom that communicate the broken electroweak symmetry to the (technicolor-singlet) standard-model fermions~\cite{Dimopoulos:1979es,Appelquist:2003hn}.\footnote{See Ref.~\cite{etcrev} for some recent reviews of technicolor and extended technicolor.} A longstanding mystery of how extended technicolor could produce suitably small neutrino masses was resolved recently~\cite{Appelquist:2002me}.

Let us now explore the \SMRn{1} \textit{Gedanken} world in greater detail. The gross properties of nucleons derived from QCD would be little changed if the $u$ and $d$ quark masses were reduced to zero. Chiral-perturbation-theory calculations suggest that the isoscalar nucleon mass would decrease from $939\mev$ to approximately $870\mev$~\cite{Gasser:1984gg,Fettes:2000xg,Beane:2002xf}. A small real-world contribution from the strange-quark sea would be absent.

The very low scale of electroweak symmetry breaking (at the scale of confinement and spontaneous chiral symmetry
breaking in QCD) brings with it a number of important consequences. In the standard electroweak theory, we \textit{choose} the value of $v = (G_{\mathrm{F}}\sqrt{2})^{-1/2} \approx 246\gev$ to reproduce the low-energy phenomenology of the charged-current weak interactions. In the \SMRn{1}, the analogue of the Fermi constant is now a \textit{prediction,} 
\begin{equation}
\overline{G}_{\mathrm{F}} = \frac{1}{\bar{f}_\pi^2\sqrt{2}} \approx 93.4\gev^{-2} \approx 8\times 10^6 \,G_{\mathrm{F}} .
\label{eq:gfbar}
\end{equation}
 As
a consequence, the strength of weak interactions in this {\it Gedanken}
world is much closer to the strength of the residual strong interactions than
in the real world. 

For processes involving momentum transfers $\abs{q}$ small compared with $\overline{M}_W,\overline{M}_Z$, cross sections and decay rates would be scaled up by the prodigious factor $(\overline{G}_{\mathrm{F}} / G_{\mathrm{F}})^2 \approx 6.4 \times 10^{13}$. This factor applies, for example, to nuclear beta decay in the \SMRn{1} \textit{Gedanken} world. Outside the regime of negligible momentum transfers, the coupling-constant amplification is tempered by the damping influence of the gauge-boson propagator.  In the case of inverse beta decay, $\nu_e n \to e p$, the scaling factor is still large enough to increase the real-world cross section (roughly $10^{-38}\cm^2$) to the level of a few millibarns.

\subsubsection{The hadron spectrum}
In addition to the gauge bosons and massless electron and neutrino, the low-energy spectrum of the \SMRn{1} contains color-singlet hadrons composed of the massless up and down quarks and gluons. 
The pions (would-be Nambu--Goldstone bosons) are absent from the physical spectrum, having been absorbed as the longitudinal components of the electroweak gauge bosons in a sort of merger of the strong and electroweak degrees of freedom. 
Small mixings may occur between the massive $W^\pm$ bosons and the charged spin-1 mesons $\rho^\pm$, $a_1^\pm$ and between the $Z^0$ boson and the neutral spin-1 mesons $\rho^0$, $\omega$, $a_1^0$.  The real-world decays of mesons into pions would be replaced in the \SMRn{1} by decays into weak gauge bosons, such as $\rho^+ \to W^+Z$, $\rho^0 \to W^+W^-$, $\omega \to W^+W^-Z$. As in the real world, the meson spectrum would include superpositions of $q\bar{q}$ and glueball states, including some that are primarily glueballs.

The chromomagnetic hyperfine interaction elevates the isospin-$\cfrac{3}{2}$, spin-$\cfrac{3}{2}$ $\Delta$ states above the spin-$\cfrac{1}{2}$ nucleon isodoublet. The real-world decays of $\Delta(1232) \to N\pi$ would be replaced in the \SMRn{1} by $\Delta \to N(W^\pm,Z,\gamma)$.

We have already remarked that the nucleon mass in the \SMRn{1} \textit{Gedanken} world with massless quarks would be decreased from its real-world value by less than 10\%, because the great bulk of the nucleon masses arises not from quark masses, but from confinement energy. The small masses of the up and down quarks are an inessential element of the big picture. An interesting---and delicate---question concerns the relative masses of the neutron and proton, for which the $u$ and $d$ quark masses do play an essential role in shaping the real world.

If electromagnetic self-energy corrections were the only relevant factor in determining the neutron-proton mass difference, it would be  intuitively obvious that the electrically charged proton should outweigh its electrically neutral partner, the neutron. However, that conclusion is false in the real world: $M_n - M_p \approx 1.293\mev$~\cite{Amsler:2008zz}. The electromagnetic contribution to the nucleon masses, larger for the proton ($uud$) than for the neutron ($udd$), is more than compensated by the fact that the current-quark mass of the down-quark exceeds that of the up: $m_d > m_u$. It is natural to guess that in the \SMRn{1}, with massless up and down quarks, the proton should outweigh the neutron. The issue is considerably more subtle, because in the \textit{Gedanken} world---in contrast to the real world---weak-interaction contributions to the $n$-$p$ mass difference are not negligible, and these tend to make the neutron outweigh the proton. We cannot make a definite statement about which nucleon is the lighter; we describe the ingredients of the analysis in Appendix~\ref{app:npdiff}.

\subsubsection{Accelerated radioactive $\beta$ decay}
Since the weak interactions are close in strength to the strong interactions in the \SMRn{1}\ \textit{Gedanken} world, the beta decay of the heavier member of the nucleon doublet is much more rapid than in the real world, for a given value of the $n$-$p$ mass difference.
The rate for neutron $\beta$-decay in the real world scales approximately as $\Gamma \propto G_{\mathrm{F}}^2\abs{M_n - M_p}^5$, so the corresponding expression for the $\beta$-decay rate of the heavier nucleon in the \SMRn{1} is $\overline{\Gamma} \propto \overline{G}_{\mathrm{F}}^2\abs{\overline{M}_p - \overline{M}_n}^5$. If as a simple illustration we take $\abs{\overline{M}_p - \overline{M}_n} =  \abs{M_n - M_p}$, then the lifetime of the heavier nucleon in the \SMRn{1} would be 
\begin{equation}
\bar{\tau} \approx \left(\frac{\bar{f}_\pi}{v}\right)^4 \tau_n \approx 15\ps\;.
\label{eq:plifetime}
\end{equation}
If the proton outweighs the neutron, then 
protons present in the early universe would decay extremely rapidly, so that the dominant species present at late times (temperature well below the QCD confinement phase transition) would be neutrons, neutrinos and antineutrinos, and a plasma of electron-positron pairs. There would be no protons, and so no possibility of hydrogen atoms. On this scenario, the lightest nucleus would be a single neutron which, being neutral, could not bind any electrons to it. 

\subsubsection{A classic unitarity bound revisited \label{subsubsec:needw}}
Partial-wave unitarity has provided important guidance about the range of applicability of low-energy effective theories and the path to more comprehensive theories. A classic example is the analysis of inverse muon decay  in the four-fermion (current-current) description of the charged-current weak interactions~\cite{Lee:1965js}. The invariant amplitude $\mathcal{M}$ for a reaction may be decomposed into partial-wave amplitudes $a_J$ according to ${\cal M}(s,t)=16\pi\sum_J(2J+1)a_J(s)P_J(\cos{\theta})$, where $s$ and $t$ are Mandelstam invariants, the $P_J$ are Legendre polynomials, and $\theta$ is the c.m.\ scattering angle. Partial-wave unitarity requires that $\abs{a_J} \leq 1$.\footnote{ Several forms of the unitarity constraint appear in the literature; $\abs{a_J} \le 1$ is the least restrictive, and the one we adopt.  The partial wave amplitude $a_J$  can be written as $a_J=(\eta_J e^{2i\delta_J}-1)/(2i)$, where $\eta_J$ measures the inelasticity and $\delta_J$ the phase shift.
A general expression of unitarity for the full amplitude is  ${\rm Im}(a_J)
\ge |a_J|^2$ (which is saturated if $\eta_J=1$ and $\delta_J=\cfrac{\pi}{2}$, for which
$a_J=i$). Using $|a_J|^2 = ({\rm Re}(a_J))^2 + ({\rm Im}(a_J))^2$, one can
derive the condition  $\abs{{\rm Re}(a_J)} \le \cfrac{1}{2}$.  However, the
tree-level Born amplitude $a_J$ is real, and hence it actually violates the lower bound on $\mathrm{Im}(a_J)$.  The different versions of the unitarity conditions reflect the fact that the lowest-order amplitude in perturbation theory is only an approximation to the exact
amplitude.}

In Fermi's current-current theory, the $s$-wave amplitude for $\nu_\mu e \to \mu \nu_e$ is
\begin{equation}
a_0 = G_{\mathrm{F}}s/\pi\sqrt{2} ,
\label{eq:pwinvmu}
\end{equation}
so partial-wave unitarity implies that the theory can only be valid for
\begin{equation}
\sqrt{s} \le \left(\frac{\pi\sqrt{2}}{G_{\mathrm{F}}}\right)^{1/2} \approx 620\gev.
\label{eq:invmumax}
\end{equation}
This is a sign that the structure of the theory must change below $620\gev$. Nature's choice, exhibited in the standard electroweak theory, is to introduce the $W$ boson, which softens the four-fermion coupling. The observed mass, $M_W \approx 80.4\gev$, shows that new physics does appear below the critical c.m.\ energy.

To make the analogous calculation in the \SMR, we must look ahead to the two-generation \textit{Gedanken} world discussed in \S\ref{subsec:moregen}, for which we find [\cf \Eqn{eq:nggf}] that $\overline{G}_{\mathrm{F}}$ scales inversely with the number of generations. The corresponding point-coupling theory could only be valid up to 
\begin{equation}
\sqrt{s} \le \left(\frac{\pi\sqrt{2}}{\overline{G}^{(n_g=2)}_{\mathrm{F}}}\right)^{1/2} \approx 310\mev.
\label{eq:smrinvmu}
\end{equation}
This bound is consistent with the appearance of the $W$ boson with mass $\overline{M}_W \approx 40\mev$ in the \SMRn{2} \textit{Gedanken} world [\cf \Eqn{eq:ngmw}].

\subsubsection{Strong scattering of electroweak gauge bosons \label{subsubsec:lqt}}
In addition to hiding the electroweak symmetry in the standard model, the Higgs boson plays an important role in regulating the high-energy behavior of scattering amplitudes. A \textit{Gedanken} experiment that illuminates that role~\cite{Lee:1977eg} leads to a conditional upper bound on the Higgs-boson mass that sets a key target for experiment in the real world. 

It is straightforward to compute the 
amplitudes ${\cal M}$ for gauge boson scattering at high energies, and to make
a partial-wave decomposition.
 Most channels ``decouple,'' in the sense 
that partial-wave amplitudes are small at all energies (except very
near the particle poles, or at exponentially large energies), for
any value of the Higgs boson mass $M_H$. Four channels are interesting: $\ket{W^+W^-}, \ket{ZZ/\sqrt{2}}, \ket{HH/\sqrt{2}}$, and $\ket{HZ}$, where all the gauge bosons are understood as the longitudinal polarization
states, and the factors of $\sqrt{2}$ account for identical particle
statistics. For these, the $s$-wave amplitudes are all asymptotically
constant (\ie, well-behaved) and  
proportional to $G_{\mathrm{F}}M_H^2.$ In the high-energy 
limit,\footnote{It is convenient to calculate these amplitudes by 
means of the Goldstone-boson equivalence theorem~\cite{Cornwall:1974km}, which 
reduces the dynamics of longitudinally polarized gauge bosons to a 
scalar field theory with interaction Lagrangian given by 
$\mathcal{L}_{\mathrm{int}} = -\lambda v h 
(2w^{+}w^{-}+z^{2}+h^{2}) - 
(\lambda/4)(2w^{+}w^{-}+z^{2}+h^{2})^{2}$, with $1/v^{2} = 
G_{\mathrm{F}}\sqrt{2}$ and $\lambda = G_{\mathrm{F}}M_{H}^{2}/\sqrt{2}$. \label{fn:rufus}}
\begin{equation}
\lim_{s\gg M_H^2}(a_0)\to\frac{-G_{\mathrm{F}} M_H^2}{4\pi\sqrt{2}}\cdot \left[
\begin{array}{cccc} 1 & 1/\sqrt{8} & 1/\sqrt{8} & 0 \\
      1/\sqrt{8} & 3/4 & 1/4 & 0 \\
      1/\sqrt{8} & 1/4 & 3/4 & 0 \\
      0 & 0 & 0 & 1/2 \end{array} \right] \; .
\end{equation} 
Requiring that the largest eigenvalue respect the 
partial-wave unitarity condition $\abs{a_0}\le 1$ yields
\begin{equation}
	M_H \le \left(\frac{8\pi\sqrt{2}}{3G_{\mathrm{F}}}\right)^{1/2} =4v\sqrt{\pi/3} =1\tev
	\label{eq:Hbound}
\end{equation}
as a condition for perturbative unitarity.

If this (standard-model) bound is respected, weak interactions remain weak at all
energies, and perturbation theory is everywhere reliable. If the
bound is violated, perturbation theory breaks down, and weak
interactions among $W^\pm$, $Z$, and $H$ become strong on the TeV scale.
This means that the features of strong interactions at GeV energies
will come to characterize electroweak gauge boson interactions at
TeV energies. 
The bound \eqn{eq:Hbound} means that new phenomena---either a Higgs boson or strong dynamics---are to
be found in real-world electroweak interactions at energies not much larger
than 1~TeV.\footnote{In the standard-model context there are also
upper bounds on (i) the quartic coupling $\lambda$ and hence the Higgs mass
\cite{higgstriv}, and (ii) Yukawa couplings and hence fermion masses generated
by Yukawa interactions \cite{yuktriv}, arising from triviality constraints.}

The models we are considering have no Higgs boson, but it is informative to write down the formal analogue of the unitarity bound \Eqn{eq:Hbound} on the Higgs-boson mass in the \SMRn{1} \textit{Gedanken} world. It is 
\begin{equation}
	\overline{M}_H \le \left(\frac{8\pi\sqrt{2}}{3\overline{G}_{\mathrm{F}}}\right)^{1/2} =4\bar{f}_\pi\sqrt{\pi/3} \approx 350\mev.
	\label{eq:SMRHbound}
\end{equation}
This condition is of course violated in the world without Higgs, which resembles the standard model in the limit of $M_H \to \infty$. We would therefore expect to see strong scattering among weak bosons as in the ordinary strong interactions among pions. 

A relevant detailed analysis was carried out in the context of the standard electroweak theory (with a heavy Higgs boson) by Chanowitz and Gaillard~\cite{Chanowitz:1985hj}. We evaluate the amplitudes for electroweak gauge-boson scattering in the limit of squared-c.m.-energy $s \gg M_W^2, M_Z^2$, with the Higgs-boson mass formally taken to infinity. In this limit, the $J=0$ partial-wave amplitudes for the normalized states $\ket{W^+W^-}$ and $\ket{ZZ/\sqrt{2}}$ are given by
\begin{equation}
a_0 = \frac{s}{32\pi v^2}\left[ \begin{array}{cc}
1 & \sqrt{2} \\
\sqrt{2} & 0
\end{array}
\right] \;,
\end{equation}
for which the larger eigenvalue is $a_0^{\mathrm{max}} = s/16\pi v^2$. The  unitarity condition $\abs{a_0} \le 1$ is saturated for $\sqrt{s^\star} = 4\sqrt{\pi}v \approx 1.74\tev$ in the standard model with a very heavy Higgs boson. In the \SMRn{1} \textit{Gedanken} world, the requirement that $\abs{a_0^{\mathrm{max}}} \le 1$ is saturated for 
\begin{equation}
\sqrt{s^\star} = 4\sqrt{\pi}\bar{f}_\pi \approx 620\mev .
\label{eq:changail}
\end{equation}

The \SMRn{1} is a theory of strongly coupled electroweak gauge bosons---as expected for a Higgsless model---that becomes strongly coupled on the hadronic scale. In the \SMR\ context, the equivalence theorem~\cite{Cornwall:1974km} noted in Footnote~\ref{fn:rufus} states that scattering among weak gauge bosons at energies $s \gg \overline{M}_{W,Z}^2$ is dominated by amplitudes for longitudinal polarization states, which are equivalent, up to terms of order $ \overline{M}_{W,Z}^2/s$, to the amplitudes for scattering among the corresponding Nambu--Goldstone bosons. Consequently, we can examine the strongly coupled gauge boson scattering in the \textit{Gedanken} world in the light of experimental and theoretical knowledge of $\pi\pi$ scattering in the real world, suitably rescaled in mass.

As in the standard model, the $(I=0,J=0)$ and $(I=1, J=1)$ channels and their Regge recurrences are attractive, while the $(I=2, J=0)$ channel is repulsive. When the partial-wave amplitudes approach their bounds, resonances will form and multiple production of gauge bosons will ensue, in emulation of phenomena seen in $\pi\pi$ scattering in the real world. The (nonperturbative) resonant contributions unitarize the partial-wave amplitudes.  

In real-world $\pi\pi$ scattering, strongly coupled resonances include the isovector vector meson $\rho(775)$, the isoscalar tensor meson $f_2(1270)$, and the isoscalar scalar meson $f_0(980)$, all of which decay dominantly into $\pi\pi$. Similar structures could be expected in the \SMRn{1}\ \textit{Gedanken} world. Theoretical tools developed to address $\pi\pi$ dynamics can be applied there~\cite{Basdevant:1969sz}. Scaling arguments derived in the context of the minimal (QCD-like) technicolor model~\cite{Dimopoulos:1979sp,Dimopoulos:1980yf} suggest that
\begin{equation}
\overline{M}_\rho = (\bar{f}_\pi/f_\pi) M_\rho \approx 730\mev,
\label{eq:mrhogw}
\end{equation}
and
\begin{equation}
\frac{\overline{\Gamma}(\rho \to W^+W^-)}{ \Gamma(\rho \to \pi^+\pi^-) } = \frac{\overline{M}_\rho}{M_\rho}\left[\frac{1 - 4\overline{M}_W^2/\overline{M}_\rho^2}{1 - 4{M}_\pi^2/{M}_\rho^2}\right]^{3/2} \!\!\approx 170\mev.
\label{eq:gamrhogw}
\end{equation}
Investigations of different unitarization schemes to the heavy-Higgs version of the standard model, which were informed by the experience of $\pi\pi$ scattering, would find ready application to the \textit{Gedanken} world~\cite{Dobado:1990jy}.

\subsubsection{Strong-interaction symmetries \label{subsubsec:sisym}}
An outstanding challenge in real-world quantum chromodynamics is to understand why \textsf{CP} and \textsf{T} are observed to be good symmetries of the strong interaction. The issue arises because of the generic presence in the QCD Lagrangian of the term,
\beq
\mathcal{L}_{\theta} = \frac{\theta g_s^2}{32 \pi^2}
G^a_{\mu\nu}\widetilde G^{a\,\mu\nu},
\label{ffdual}
\eeq
where $g_s$ is the strong coupling, $G^a_{\mu\nu}$ is the gluon field-strength tensor with color-index $a$, $\widetilde G^{a\,\mu\nu}$ is its dual, and the implied sum over $a$ runs over the adjoint representation of the color gauge group. Because of the nontrivial topology of the color gauge fields, it is not possible to remove $\mathcal{L}_{\theta}$ as a total derivative. If one or more of the quarks has zero hard mass, the argument of the quark mass matrix becomes a free parameter, and the physical value of the $\theta$-parameter can be tuned to zero.   In the  \SMR, all of the fermions have zero bare masses, so there is no strong \textsf{CP} problem due to instantons.
There is strong evidence that all the quark masses (and, in particular, the up-quark mass) are nonzero in the real world~\cite{qmmini}. 

In the real world, color-singlet hadrons experience residual strong interactions
that can be modelled at long distances ($\gtrsim 1\fm$) by one-pion
exchange, and at shorter distances by the exchange of $\rho$ and $\omega$ vector mesons and effects due to overlapping quark wavefunctions, which yield a repulsive hard-core component to
the nucleon-nucleon interaction.  The \SMRn{1} \textit{Gedanken} world lacks physical pions, because the corresponding degrees of freedom have become the longitudinal components of $W^\pm$ and $Z^0$. The long-range part of the $NN$ interaction is supplied by the exchange of the electroweak gauge bosons themselves, because of their small masses and the enhanced strength of the weak interactions in the \SMR. The consequence for strong-interaction symmetries is dramatic: in the  \textit{Gedanken} world, parity (\textsf{P}) and charge conjugation (\textsf{C}) are violated, because the long-range ``strong'' force is derived from the \ewgg\ chiral gauge theory!

\subsubsection{Quantitative changes to the nuclear force \label{subsubsec:nucforce}}
Calculations of nuclear properties have reached an advanced---and, overall, very satisfying---state, through the systematic application of semi-empirical nuclear potential models guided by fundamental principles~\cite{npbook}. It would be excessively ambitious to attempt to characterize the nature of binding among nucleons in any variant of the \SMR. A few simple observations are revealing.

As we have just noted, the long-range force of the two-nucleon interaction is governed in the \SMRn{1} by the exchange of electroweak gauge bosons. The virtual emission and reabsorption of $W^\pm$ and $Z^0$ is the analogue in this \textit{Gedanken} world of the pion cloud in the real world. Accordingly, the size of a hadron, measured in the real world by $r \simeq 1/m_\pi \approx 1.4\fm$, is given in the \SMRn{1} by $\bar{r} \simeq 1/\overline{M}_{W,Z} \approx 7\fm$, about five times greater.

In the real world, of course, the one-pion-exchange amplitude between nucleons 
involves strong couplings with strength given by $g_{\pi NN} \approx 14$ at each $\pi NN$ vertex. In the limit of low momentum transfer,
\beq
A(N_1 N_2 \to N_3 N_4) \sim {\frac{g_{\pi NN}^2}{m_\pi^2}}\;.
\label{amprw}
\eeq
The corresponding $W$-exchange amplitude in the \SMRn{1} is
\beq
\overline{A}(N_1 N_2 \to N_3 N_4) \sim \frac{(g^2)}{8\overline{M}_W^2} \sim  \frac{1}{2\bar{f}_\pi^2}\;. 
\label{amp}
\eeq
Hence, the ratio of the \textit{Gedanken}-world amplitude to the real-world amplitude is
\begin{equation}
\bar{A}/A = \frac{m_\pi^2}{2\bar{f}_\pi^2 g_{\pi NN}^2} = 0.0065\;.
\label{ampratio}
\end{equation}
Thus, although the range of the $NN$ interaction is considerably longer in the \SMRn{1} than in the real world, the strength is considerably weaker, at zero momentum transfer. However, this does not mean that a bound-state spectrum would be entirely absent.

A reasonable simplified description of $NN$ binding in the real world can be had by solving the Schr\"{o}dinger equation in a square-well potential with a radial size $a \approx 1.4\fm$ and depth $V_0$. The  occurrence of bound states is the determined by the value of the dimensionless parameter,
\beq
\xi = \frac{2\mu V_0 a^2}{\hbar^2 \pi^2} ,
\label{xi}
\eeq
where $\mu$ is the reduced mass, equal to $M_N/2$ for the two-nucleon problem.  When $\xi$ exceeds a minimal number of order unity, a first bound state appears, and as $\xi$
increases, more bound states appear in the discrete spectrum of the
Hamiltonian. The relevant figure of merit is $V_0 a^2$.  

To compare one-pion exchange in the real world with electroweak-gauge-boson exchange in the \SMRn{1}, we measure $V_0$ by the strength of the amplitude $A$ and set $a$ equal to the Yukawian range of the force. The ratio
\begin{equation}
\bar{\xi}/\xi = \frac{m_\pi^2}{2\bar{f}_\pi^2 g_{\pi NN}^2}\cdot \frac{m_\pi^2}{\overline{M}_W^2} \approx 1/6\;
\label{eq:xirat}
\end{equation}
is smaller, but not grossly smaller, than unity. This observation would have to inform studies of nucleosynthesis in the \textit{Gedanken} world.

\subsubsection{Weak instantons (Sphalerons) \label{subsubsec:sphal}}
In the real world, weak \wigg\  instantons, which violate $B$ and $L$,
conserving $B-L$, have a negligibly small effect at temperatures  $T$ much lower
than the electroweak scale $v \approx 246\gev$, because their contributions to physical
processes are suppressed by the factor $\exp(-8 \pi^2/g^2)$ at zero temperature, where $g$ is the \wigg\ gauge coupling~\cite{'tHooft:1976up}.  For $T \gtrsim v$, the sphalerons tend to erase a pre-existing baryon asymmetry of the universe, and could under some conditions generate a significant baryon asymmetry~\cite{Kuzmin:1985mm,Cohen:1993nk,Balazs:2007pf}.
The analogous
statement is true in the \SMRn{1}\ \textit{Gedanken} world, with the critical temperature rescaled from $v \to \bar{f}_\pi$.

\subsubsection{(Nearly) massless fermions}
We have emphasized that in the \SMR, leptons are massless, even after the gauge symmetry is broken as a consequence of the spontaneous breaking of the chiral symmetry of massless quarks within QCD. Zero or extremely small masses for the neutrinos do not have any negative consequences. However, the masslessness of the electron in the \SMRn{1} undermines the integrity of matter, and also makes the \SMR\ vacuum unstable against the spontaneous production of charged-lepton pairs.  The presence
of even very small electromagnetic fields with $F_{\mu\nu}F^{\mu\nu}=2(|\mathbf B|^2-|\mathbf E|^2) < 0$, for which one could transform to a Lorentz frame in which ${\mathbf B}=0$, would lead, through the
Schwinger mechanism \cite{Schwinger:1951nm}, to decay by production of $e^+e^-$ pairs. The rate for this production is sizeable for $e|{\mathbf E}| \gtrsim m_e^2$. 

In the real world, a recombination of free electrons and protons takes place when the age of the universe is about $380\,000$ years, when the mean temperature is a fraction of an electron volt. It is hard to imagine anything of the sort occurring in the $e^+e^-$ plasma of the \textit{Gedanken} world. 
Moreover, the presence of zero-mass unconfined charged particles leads to a pathology associated with the infinite acceleration to which they would be subjected under the influence of an electric field~\cite{Case:1962zz}.

If we regard the \SMR\ as a low-energy effective theory, then an ultraviolet completion will in general induce nonzero fermion masses. Consider a minimalist scenario in which the \SMRn{1} is embedded in a unified theory governed at high energies by some simple gauge group. Assume that at the unification scale $M_{\mathrm{U}}$, the common value of the gauge couplings is $\alpha_{\mathrm{U}}$. The extended quark-lepton families in a unified theory mean that proton-positron mixing is a logical consequence. 

A tree-level process that contributes to proton decay is $u + u \to X^{4/3} \to e^+ d^c$, where $d^c$ corresponds to the $d$ antiquark and $X^{4/3}$ is a leptoquark gauge boson.  At energies very low compared with $M_X \approx M_{\mathrm{U}}$, we can estimate the matrix element of the $uude$ four-fermion operator between $\ket{p}$ and $\ket{e^+}$ as
\begin{equation}
\varepsilon \equiv \mathcal{M}(p \leftrightarrow e^+) \approx \frac{4\pi\alpha_{\mathrm{U}}}{M_X^2}\Lambda_{\mathrm{QCD}}^3\;.
\label{amppeplus}
\end{equation}
It follows that the $(e^+, p)$ mass matrix is given by
\beq
\mathsf{M} = \left( \begin{array}{cc}
    0         & \varepsilon   \\
    \varepsilon ^*  & M_p  \end{array} \right )
\label{M1}
\eeq

Diagonalizing this matrix, we find a negligibly small shift in the proton
mass, a tiny $\varepsilon/M_X$ positron admixture in the proton wave function (and vice versa), and a nonzero but exquisitely small positron mass given (after a field redefinition to eliminate a physically insignigicant minus sign) by a seesaw-type formula,
\beq
m_e = \frac{\abs{\varepsilon}^2}{M_p} \;.
\label{me}
\eeq
With illustrative values $1/\alpha_{\mathrm{U}} \approx 24$ and 
$M_X \approx 10^{17}\gev$, we find that $| \varepsilon | \approx 10^{-36}\gev$ and so
$m_e \approx 10^{-72}\gev$. Independent of the infinitesimal result, it is fascinating to contemplate the possibility of mass eigenstates that mix elementary and composite fermions.

Analogous arguments apply to neutron--(anti)neutrino mixing and a similar induced mass for the neutrino.
Richer ultraviolet completions that populate lower mass scales could well lead to larger induced masses for the neutrino and electron.

\subsubsection{{A variant: $\mathrm{SU(2)_L}$ dominant \label{subsubsec:weakdom}}}
We chose to investigate \textit{Gedanken} worlds in which the hierarchy of the gauge interactions is like that of the real world, with QCD the strongest of the interactions at low energies. This protocol is sufficient for our purpose of inquiring what the world would be like in the absence of a mechanism introduced to hide the electroweak symmetry. Other deviations could be considered, of course. One that immediately suggests itself is to ask what the outcome would be if, in a world without a Higgs mechanism, the \wigg\ interaction were stronger than the \cgg\ interaction. 

The consequences are very dramatic: if the charged-current weak interaction, which is asymptotically free, is the first to become strong, it produces \wigg-singlet condensates involving two quark fields, or two lepton fields, or one quark field and one lepton field. These condensates break the color gauge group $\cgg \to \mathrm{SU(2)_c}$, break weak hypercharge and electric charge, and break baryon number and lepton number! At a lower scale, the residual color interaction confines colored objects into mesons, bosonic baryons, and glueballs. This \textit{Gedanken} world has so little in common with the real world that we shall not pursue it further.

\subsection{Multiple massless generations \label{subsec:moregen}}
With the one-generation toy model as background, we now turn our attention to how the three-generation real world would be changed if there were no Higgs mechanism. The results are similar in character for $n_g \ge 2$ fermion generations, subject to the constraint that $n_g$ is not so large as to change the qualitative character of the QCD sector.  It is well known that QCD remains asymptotically free so long as $(33 - 4n_g) > 0$, assuming $2n_g$ active quark flavors, so that up to eight generations are permitted. For our purposes, we must be more demanding and require that, as in the real world, QCD confine and spontaneously break chiral symmetry. 
\subsubsection{Electroweak symmetry breaking}

With $n_g$ fermion generations,  the modified standard model \SMRn{n_g} displays a global \chiral{2n_g}\ chiral symmetry, reflecting the presence of $2n_g$ massless quarks.\footnote{The global U(1)$_{\mathrm{A}}$ symmetry is broken by \cgg\  instantons.} So long as $n_g \lesssim 6$, QCD confines the massless quarks at an energy scale that, for given $n_g$, we denote as $\Lambda_{\mathrm{QCD}}$. It forms color-singlet $\langle \bar q q \rangle$ condensates that spontaneously break \chiral{2n_g} to an $\mathrm{SU}(2n_g)_{\mathrm{V}}$ flavor symmetry that is exact in the absence of electroweak interactions~\cite{Appelquist:1986an, Appelquist:1988yc, Appelquist:2007hu}.

Leaving implicit the trace over colors to make color singlets, the generalized pions are
\begin{eqnarray}
\ket{\Pi^+} & = & \frac{1}{\sqrt{n_g}}\sum_{i = 1}^{n_g}\ket{u_i\bar{d}_i} \nonumber \\
\ket{\Pi^0} & = &  \frac{1}{\sqrt{2n_g}}\sum_{i = 1}^{n_g}\ket{(u_i\bar{u}_i - d_i\bar{d}_i)} \\
\ket{\Pi^-} & = & \frac{1}{\sqrt{n_g}}\sum_{i = 1}^{n_g}\ket{d_i\bar{u}_i}. \nonumber 
\label{eq:genpip}
\end{eqnarray}
These are three of the $4n_g^2 -1$ Nambu--Goldstone bosons\footnote{The flavor-singlet pseudoscalar meson, the generalized $\eta'$ corresponding
to the $\mathrm{U(1)_A}$ generator is massive, since this symmetry is anomalous.}
 of the \SMRn{n_g}, before the \ewgg\ interactions are turned on.

As before, we treat the \ewgg\ interactions as a perturbation on the situation created by QCD. The $\Pi^\pm$ and $\Pi^0$ are absorbed as the longitudinal components of the massive $W^\pm$ and $Z^0$, for which the masses are now given as 
\begin{equation}
\overline{M}_W^2 = n_g\,g^2\bar{f}_\pi^2/4
\label{eq:ngmw}
\end{equation}
 and 
 \begin{equation}
 \overline{M}_Z^2 = n_g(g^2 + g^{\prime\,2})\bar{f}_\pi^2/4 ,
 \label{eq:ngmz}
 \end{equation}
  so that in the \SMRn{n_g},
\begin{equation}
\overline{M}_W \approx 28\sqrt{n_g}\mev; \qquad \overline{M}_Z \approx 32\sqrt{n_g}\mev .
\label{eq:mwng}
\end{equation}
The custodial-symmetry relation \eqn{eq:wzrat} remains in force.
In the \SMRn{n_g} \textit{Gedanken} world, the counterpart of the real-world Fermi constant is 
\begin{equation}
\overline{G}_{\mathrm{F}} = \frac{1}{n_g\bar{f}_\pi^2\sqrt{2}}= G_{\mathrm{F}} \cdot \frac{v^2}{n_g\bar{f}_\pi^2}.
\label{eq:nggf}
\end{equation}

The scaled-down value of $\overline{G}_{\mathrm{F}} \propto 1/n_g$ relative to what we found in the \SMR\ rescales the constraints derived from partial-wave unitarity in \S\ref{subsubsec:lqt}. In the more general \SMRn{n_g}, the tipping point \eqn{eq:SMRHbound} for the mass of the (nonexistent) Higgs boson would come for
\begin{equation}
	\overline{M}_H \le \left(\frac{8\pi\sqrt{2}}{3\overline{G}_{\mathrm{F}}}\right)^{1/2} =\frac{4\bar{f}_\pi\sqrt{\pi/3}}{\sqrt{n_g}} \approx \frac{360\mev}{\sqrt{n_g}}.
	\label{eq:SMRHboundng}
\end{equation}
Of greater relevance, in the \SMRn{n_g} \textit{Gedanken} world, the electroweak-gauge-boson scattering amplitudes saturate partial-wave unitarity, in the sense that $\abs{a_0^{\mathrm{max}}} = 1$, for (\cf \eqn{eq:changail})
\begin{equation}
\sqrt{s^\star} = 4\sqrt{\pi n_g}\bar{f}_\pi \approx 620\,\sqrt{n_g}\mev. 
\label{eq:changailng}
\end{equation}

\subsubsection{The hadron spectrum}
Let us consider the fate of the $(4n_g^2 -1)$ Nambu--Goldstone bosons of the \SMRn{n_g} \textit{Gedanken} world in greater detail. There are $n_g^2$ NGBs with charge $+ 1$ and an equal number with charge $-1$. These are the counterparts of the real-world $\pi^\pm$ for $n_g = 1$, plus $K^\pm$, $D^\pm$, and $D_s^\pm$ for $n_g = 2$, etc. Next, there are $2n_g(n_g-1)$ electrically neutral NGBs that carry nonvanishing flavor quantum numbers. No such particles exist in the real world with one generation; for $n_g =2$, the corresponding particles are $K^0, \bar{K}^0$, and $D^0, \bar{D}^0$. Finally, there are $2n_g - 1$ self-conjugate neutral, but flavor-nonsinglet, particles. The real-world analogues are $\pi^0$ at one generation, plus $\eta$ and $\eta_c$ in the second generation. 

When the \ewgg\ interactions are turned on, the generalized pions disappear from the hadron spectrum, leaving $(4n_g^2 - 4)$ NGBs.\footnote{The persistence of $n_g^2-1$ massless charge $+1$ bosons and their oppositely charged partners constitutes a new source of vacuum instability not present in the \SMRn{1}.} That is to say, the \SMRn{n_g \ge 2} \textit{Gedanken} worlds exhibit  massless (or very light) strongly interacting Nambu-Goldstone bosons. To estimate the coupling of the NGBs to baryons, we note that an analogue of the Goldberger-Treiman relation~\cite{Goldberger:1958tr} 
\beq
\abs{g_A} M_N ={f}_\pi g_{\pi NN} ,
\label{gtrel}
\eeq
where $g_A = -1.2695 \pm 0.0029$ is the axial charge of the neutron~\cite{Amsler:2008zz}, should hold in the \SMR. This equality holds within 15\% in the real world, with $g_{\pi NN} \approx 14$. In the \SMR\ \textit{Gedanken} world, we have noted the estimates that the pion decay constant and the nucleon mass would be slightly reduced with respect to their real-world values, and we expect little change in the axial charge $g_A$. We can therefore infer that the \SMR-value of $\bar{g}_{\pi NN} \approx g_{\pi NN}$ in the real world. The $\mathrm{SU}(2n_g)$ flavor symmetry implies that this coupling strength applies to all the NGB--baryon interactions. 

The size of hadrons is determined by the radius of the NGB cloud, which would be far greater than the $1.4\fm$ size of the pion cloud in the real world. Similarly, the range of the baryon-baryon interaction would be very great. In practice, this would mean that the wavefunctions of
different baryons would overlap. This, in turn, could lead to pairing effects
that would produce integral-spin dibaryon correlations, perhaps analogous to
Cooper pairs in a superconductor, with consequent macroscopic coherent quantum
effects, including, perhaps, macroscopic occupation numbers of ground states when the nucleon number density is sufficiently high and the temperature sufficiently low.

The structure of the baryon spectrum in the \SMR\ is familiar from what we observe in the real world, with one important modification. In the absence of quark mass differences, all states in a representation are highly degenerate and consequently the $\mathrm{SU}(n_q)$ flavor symmetry, with $n_q = 2n_g$, is essentially exact. In the context of QCD based on the \cgg\ gauge group, the baryons are composed of three quarks. They can be classified according to the Clebsch-Gordan
decomposition of the threefold product of $\mathbf{n_q}$, the fundamental
representation of SU($n_q$), 
\beq
\fund \otimes \fund \otimes \fund = S_3 \oplus M_1 \oplus M_2  \oplus A_3
\label{baryon}
\eeq
where $A_3$, $S_3$, and $M$ denote the totally antisymmetric and symmetric 
rank-3 tensor representations, and $M$ has mixed symmetry.  These
have dimensions
\begin{eqnarray}
{\rm dim}(S_3)&  = & \frac{n_q(n_q+1)(n_q+2)}{3!} \ , \nonumber \\
{\rm dim}(M) & = & \frac{n_q(n_q^2-1)}{3} \ , \\
{\rm dim}(A_3) & = & {n_q \choose 3} \ . \nonumber
\label{eq:repdims}
\end{eqnarray}

In the real world, the lowest-lying states are the spin-$\cfrac{1}{2}$ nucleon doublet, baryon octet, etc., for $n_q = 2, 3, \ldots$, corresponding to the  mixed representation, and the spin-$\cfrac{3}{2}$ $\Delta$ quartet, baryon decimet, etc., corresponding to the symmetric representation. As in the \SMRn{1}\ case, chromomagnetic interactions would raise the spin-$\cfrac{3}{2}$ states above the spin-$\cfrac{1}{2}$ ground state by an amount of order $\Lambda_{\mathrm{QCD}}$.

\section{Including explicit \newline fermion masses \label{sec:baremass}}
As we have noted, the  \SMR\ theory in isolation has an infrared pathology
resulting from the massless unconfined charged leptons.  Even if one takes into
account a nonzero mass that is generated in a grand-unified ultraviolet completion of the
 \SMR\ theory, this is still so small that the late-time, low-temperature
universe would still be in the radiation-dominated era with a plasma of
charged lepton pairs.  Additional problems arise with the presence of massless charged scalars in multigeneration versions of the \SMR.
The instabilities that beset the  \SMR\ with zero fermion bare masses motivate us
to investigate the properties of the corresponding modified standard model with bare fermion masses, which we designate \SMRm. It is free of the infrared pathology and
provides a different interesting theoretical laboratory in which to study the behavior
of scattering cross sections in a theory without a fundamental Higgs field. 

We add to the  low-energy effective
Lagrangian explicit fermion mass terms that appear hard (persist) on scales well above $\Lambda_{\mathrm{QCD}}$.\footnote{What here appear as hard masses could be seen as soft (induced by some spontaneous symmetry breaking) on substantially higher scales. Provided they are dynamically generated at some high
scale $\Lambda^\star \gg \Lambda_{\mathrm{QCD}}$ in the ultraviolet-complete theory, the associated running fermion masses will fall off as $m_{f} \propto \Lambda^{\star\,3}/p^2$ for Euclidean
momenta $p \gg \Lambda^\star$ \cite{Christensen:2005hm}.} Mass terms for up quarks, down quarks, and charged leptons are of the Dirac form
\begin{eqnarray}
\mathcal{L}_m	& = & -\sum_{n,n^\prime = 1}^{n_g} \left[ \bar{u}_{n\mathrm{L}}\mathbf{U}_{nn^\prime}u_{n^\prime\mathrm{R}} + \bar{d}_{n\mathrm{L}}\mathbf{D}_{nn^\prime}d_{n^\prime\mathrm{R}}
\right.
\nonumber \\
 & + & \left. \bar{\ell}_{n\mathrm{L}}\mathbf{L}_{nn^\prime}\ell_{n^\prime\mathrm{R}} \right] + \hbox{h.c.}
 \label{eq:bare}
 \end{eqnarray}
 where $\mathbf{U}, \mathbf{D}, \mathbf{L}$ are fermion mixing matrices and
 $n_g$ is the number of fermion generations. (Dirac and Majorana mass terms can be written for neutrinos as well.) By bringing the $n_g \times n_g$ mass matrices to diagonal form, one obtains the mass eigenstates and the mixing matrices for quarks and leptons.
 
 Explicit Dirac mass terms link the left-handed and right-handed fermions, and thus violate the \ewgg\ gauge symmetry of the electroweak theory. Accordingly, at lowest order in perturbation theory, scattering amplitudes for the production of pairs of longitudinally polarized gauge bosons in fermion-antifermion annihilations grow with c.m.\ energy roughly as $G_\mathrm{F}m_f E_{\mathrm{cm}}$. For a fermion with bare mass $m_f$, the resulting partial-wave amplitudes saturate partial-wave unitarity for the standard model with a Higgs mechanism at a critical c.m.\ energy~\cite{Appelquist:1987cf, Maltoni:2001dc,Dicus:2004rg},
\beq
\sqrt{s_f} \simeq \frac{4\pi\sqrt{2}}{\sqrt{3\eta_f} \, G_\mathrm{F} m_f} = 
                \frac{8 \pi v^2}{\sqrt{3\eta_f} \, m_f}\;,
\label{unitaritybound}
\eeq
where  $\eta_f=1(N_{\mathrm{c}})$ for leptons (quarks). 
As usual, the parameter $v$ sets the scale of electroweak symmetry breaking. If the electron mass were hard, the critical energy would be $\sqrt{s_e} \approx 1.7 \times 10^9\gev$; the corresponding energy for the top quark is $\sqrt{s_t} \approx 3\tev$. The fact that a hard electron mass would only imply a saturation of partial-wave unitarity at a prodigiously high energy means that the celebrated gauge cancellation observed in the reaction $e^+ e^- \to W^+ W^-$ at LEP~\cite{lepewwgww} validates the gauge symmetry of the electroweak theory, but does not establish that the theory is renormalizable~\cite{Christensen:2005hm,Appelquist:1987cf}. 

In the \SMRm, electroweak symmetry breaking is induced by QCD at the QCD scale. If we choose the up- and down-quark masses equal to the real-world values, then the appropriate scale should be close to the real-world value of $f_\pi$. Consequently, perturbative unitarity would be violated by bare fermion mass terms at a critical c.m.\
energy 
\beq 
\sqrt{s_f}  \simeq \frac{8 \pi n_g{f}_\pi^2 }{\sqrt{3\eta_f} \, m_f}\;:
\label{ourbound}
\eeq
the greater the bare fermion mass, the lower the critical energy. 

If the \SMRm\ is to be a useful construct, it must make sense up to
energies above the scale of low-lying mesons, baryons, and glueballs, 
say, up to $\sqrt{s_f} = 50{f}_\pi \approx 4.6\gev$. That can be achieved provided that 
\begin{equation}
m_f \lesssim  \frac{4\pi n_g f_\pi}{25 \sqrt{3\eta_f}} \approx \left\{
\begin{array}{l}
27\,n_g\mev\hbox{ (leptons)} \\
15\,n_g\mev\hbox{ (quarks)}\;,
\end{array}
\right.
\label{mfmax}
\end{equation}
conditions that would generously accommodate the real-world values of the $u$, $d$, and $e$ masses.

If we impose the more modest requirement that the fermion sector of the \SMRm\ respect unitarity up to the energy at which the gauge-boson scattering saturates partial-wave unitarity, $\sqrt{s^\star} = 4\sqrt{\pi n_g}{f}_\pi \approx 655\,\sqrt{n_g}\mev$ (\cf \eqn{eq:changailng}), then the requirements on bare fermion masses relax to
\begin{equation}
m_f \lesssim  \frac{2\sqrt{\pi n_g} {f}_\pi}{\sqrt{3\eta_f}} \approx \left\{
\begin{array}{l}
190\,\sqrt{n_g}\mev\hbox{ (leptons)} \\
110\,\sqrt{n_g}\mev\hbox{ (quarks)}\;.
\end{array}
\right.
\label{mfmaxU}
\end{equation}

Not only do the hard fermion masses implied by (\ref{eq:bare}) explicitly break the \ewgg\ gauge symmetry, they also break the \chiral{2n_g}\ chiral symmetry that attends massless fermions. Just as in Nambu's classic construction~\cite{Nambu:1960xd} of light, but not massless, pions by the spontaneous breaking of an approximate chiral symmetry, the QCD-induced spontaneous breaking of the \ewgg\ symmetry does not yield exactly massless spin-zero particles, but light pseudo-Nambu--Goldstone bosons (pNGBs). We recall that in the Nambu--Jona-Lasinio example~\cite{Nambu:1961fr} a bare nucleon mass of a few MeV sufficed to generate a realistic pion mass. Unlike their massless counterparts in the \SMRn{n_g>1}\ \textit{Gedanken} world, the charged pNGBs of \SMRm\ do not create a vacuum instability problem. 

As is the case for the standard model, the electroweak sector of the \SMRm\ is \textsf{CP}-conserving for one and two fermion generations. For $n_g \ge 3$, the $(n_g - 1)(n_g - 2)/2$ unremovable phases in the quark mixing matrix for the charged-current interactions provides a mechanism for \textsf{CP} violation~\cite{Kobayashi:1973fv}.

\section{Beyond three colors \label{sec:colors} }
In the modified standard models we have considered, the scale of electroweak symmetry breaking is linked directly to the confinement scale of quantum chromodynamics. This circumstance makes it interesting to examine generalizations of the \SMR\ in which the number of colors, $N_{\mathrm{c}}$, differs from the real-world value of 3.\footnote{To build specific models, we will confine our attention to odd values of $N_{\mathrm{c}}$. This restriction maintains baryons as fermions, since the simplest body plan for a multiquark color singlet is $q^{N_{\mathrm{c}}}$. A second motivation for keeping $N_{\mathrm{c}}$ odd is the requirement that the theory should be free of a
global Witten $\pi_4$ anomaly~\cite{Witten:1982fp} for arbitrary values of the number of fermion generations, $n_g$, lest the \wigg\ gauge theory be mathematically inconsistent. This means that the number of chiral components of  left-handed doublets [$=(N_{\mathrm{c}}+1)n_g$] must be even. Hence, for general values of $n_g$, $N_{\mathrm{c}}$ must be odd.}

The large-$N_{\mathrm{c}}$ limit holds particular interest, because it has yielded a number of insights into the structure of  QCD in two and four dimensions~\cite{'tHooft:1973jz,Witten:1979kh}. This limit is
\beq
N_{\mathrm{c}} \to \infty \ , \quad {\rm with} \ \ g_3^2 \, N_{\mathrm{c}} = \hbox{ finite constant} \ne 0,
\label{nclim}
\eeq
so that the strong coupling $g_3
\to 0$ as $N_{\mathrm{c}}^{-1/2}$.  

The large-$N_{\mathrm{c}}$ limit was originally applied to QCD
in isolation, neglecting the electroweak interactions.  The requirement that the \ewgg\ gauge interactions be free of anomalies imposes restrictions on the electric charges of the fermions in the theory.
Expressing the quark charges $Q_u$ and $Q_d$ in terms of the electron charge $Q_e$, we find~\cite{Shrock:1995bp}
\beq
Q_u = Q_d + 1 = \cfrac{1}{2}\left[ 1 - (2Q_e +1)/N_{\mathrm{c}} \right]
\label{qusol}
\eeq
With the canonical choice $Q_e = -1$ (for which $Q_\nu = 0$), the condition becomes~\cite{Abbas:1990kd}
\beq
Q_u = Q_d + 1 = \cfrac{1}{2}\left(  1 +{1}/{N_{\mathrm{c}}} \right),
\label{qusolusual}
\eeq
which reproduces the familiar assignment $Q_u = +\cfrac{2}{3}$ for $N_{\mathrm{c}}=3$.

Once the electroweak interactions are included, the electroweak coupling constants must be rescaled in the manner of \Eqn{nclim} to compensate for the growing number of chiral fermion degrees of freedom:
\beq
g^2 N_{\mathrm{c}} \ , \; g^{\prime\,2}N_{\mathrm{c}}, \; e^2 N_{\mathrm{c}} \to \hbox{constants as }N_{\mathrm{c}} \to \infty.
\label{gjlim}
\eeq
Without this rescaling, the theory would have unphysical properties in the $N_{\mathrm{c}}\to\infty$ limit: (i) the \ewgg\ interactions
would become infinitely strong relative to the \cgg\ interaction and (ii)
various qualitative properties, such as the stability
of mesons, would be spoiled. To cite one example, the decay rate $\Gamma(\pi^0 \to \gamma\gamma) \propto \alpha(Q_u^2-Q_d^2)N_{\mathrm{c}}$ would grow unacceptably, if $\alpha = e^2/4\pi$ were independent of $N_{\mathrm{c}}$.

The scaling of the electroweak gauge couplings (in particular $g$) compensates for the growth $f_\pi \propto \sqrt{N_{\mathrm{c}}}$ of the pion decay constant, which we recall is defined through the matrix element \Eqn{eq:fpidef} by which the charged weak current connects a generalized pion to the vacuum in leptonic decays. The foregoing discussion applies to the $N_{\mathrm{c}} \ne 3$ generalizations of both the standard model (with spontaneous \ewgg\ symmetry breaking induced by the Higgs sector) and of the \SMR.

The same interplay between the $f_\pi \propto \sqrt{N_{\mathrm{c}}}$ increase and the $g, g^\prime \propto 1/\sqrt{N_{\mathrm{c}}}$ decrease makes the $W^\pm$ and $Z^0$ masses independent of $N_{\mathrm{c}}$ in the \SMR\ (\cf \S\ref{subsec:onegen}).  The Fermi constant is related to the \wigg\ gauge coupling and the $W$-boson mass as $G_{\mathrm{F}} = g^2/(4\sqrt{2}\,M_W^2)$. If we regard the value of $G_{\mathrm{F}} = 1/(v^2\sqrt{2})$ as a constant of Nature within the framework of the standard model, then the scaling law \eqn{gjlim} means that $M_W \propto 1/\sqrt{N_{\mathrm{c}}}$. In contrast, $\overline{M}_W$ is independent of $N_{\mathrm{c}}$ in the \SMR, so $\overline{G}_{\mathrm{F}} \propto 1/{N_{\mathrm{c}}}$.

For odd values of $N_{\mathrm{c}} > 3$, the spectrum of low-lying hadrons in  the \SMR\ would consist of $(q^{N_{\mathrm{c}}}$) baryons~\cite{Dashen:1994qi} with spins \cfrac{1}{2}, \cfrac{3}{2}, \ldots \cfrac{N_{\mathrm{c}}}{2}, plus glueballs and $(q\bar{q})$ states that include $(4n_g - 4)$ Nambu--Goldstone bosons (pNGBs in the \SMRm) and vector mesons including $W^\pm$ and $Z$. The baryons lie in $\mathrm{SU}(n_q)_{\mathrm{flavor}}$ representations contained in the product $\left(\otimes \mathbf{n_q}\right)^{N_{\mathrm{c}}}$, and have 
masses $\propto N_{\mathrm{c}}$.

\section{Left-Right--Symmetric \newline\textit{Gedanken} World \label{sec:lrsym}}
Up to this point, we have considered modified versions of the standard model based on \smgg\ gauge symmetry, and investigated how different the world would be if electroweak symmetry breaking were accomplished not by a special agent, but as an induced effect of the spontaneous breaking of chiral symmetry in QCD. Since the standard model may well emerge as the low-energy limit of a more comprehensive theory, it is worth examining where some theories beyond the standard model would lead if the Higgs mechanism were absent. As before, our aim is not to find alternative paths to the real world, but to explore how the world might have been different.

We consider first a left-right--symmetric extension of the standard-model gauge group~\cite{Mohapatra:1974hk}, 
\begin{equation}
\lrgg\;,
\label{eq:lrgg}
\end{equation}
for which QCD will once again be the agent that induces electroweak symmetry breaking. It is a characteristic of such models that the observed maximal parity violation in the weak interactions does not reflect a fundamental asymmetry of the laws of nature, but a circumstance based on the pattern of electroweak symmetry breaking. To respect real-world limits on right-handed charged currents and a $B-L$ gauge force, the $\mathrm{SU(2)_R \otimes \blgg}$ symmetry must be broken to $\mathrm{U(1)}_Y$ at a scale $\gtrsim 1\tev$. This means that the right-handed $W^\pm$-bosons must acquire a much higher mass than the known left-handed $W^\pm$, so the low-energy charged-current interaction would be left-handed to the high accuracy established experimentally. Furthermore, the mixing among the neutral-gauge--boson eigenstates must be such as to reproduce the neutral-current properties of the standard model to high precision. 

An appealing feature of the left-right--symmetric theory is that the electric charge operator takes an elegant form in terms of generators of both of the weak-isospin gauge groups and the fundamental quantities $B$ and $L$, namely
\begin{equation}
Q = I_{3\mathrm{L}} + I_{3\mathrm{R}} + \cfrac{1}{2}(B-L).
\label{eq:qoplrdef}
\end{equation}
What the standard model identifies as weak hypercharge is given, in the \lrgg\ world, by $Y = I_{3\mathrm{R}} -\cfrac{1}{2}(B - L)$, an expression that makes sense of the curious weak hypercharge assignments [\cf\ the discussion around \Eqn{eq:chiralq} and \Eqn{eq:chirall}] in the \ewgg\ electroweak theory.

To be specific, let us consider a left-right symmetric \textit{Gedanken} world with a single generation of massless fermions and no scalar sector. We shall see presently that this \LR\ model exhibits a number of properties that differ in interesting ways from the \SMR\ models discussed in \S\ref{sec:SMR} and \ref{sec:baremass}.\footnote{It is straightforward to extend this model, as we have done for the \SMR\ models, to several fermion generations, and to consider the large-$N_{\mathrm{c}}$ limit or to admit bare fermion masses.} Now the  chiral quark fields are
\begin{equation}
\mathsf{Q}^a_{\mathrm{L}} = \left(\begin{array}{c} u^{a} \\ d^a\end{array}\right)_{\!\mathrm{L}}, \qquad \mathsf{Q}^a_{\mathrm{R}} = \left(\begin{array}{c} u^{a} \\ d^a\end{array}\right)_{\!\mathrm{R}}\;,
\label{eq:lrchiralq}
\end{equation}
with $(\cgg,\wigg,\mathrm{SU(2)_R})_{B-L}$ quantum numbers $(\mathbf{3},\mathbf{2},\mathbf{1})_{1/3}$ and $(\mathbf{3},\mathbf{1},\mathbf{2})_{1/3}$, respectively, and color index $a = 1,2,3$. The chiral lepton fields are 
\begin{equation}
\mathsf{L}_{\mathrm{L}} = \left(\begin{array}{c} \nu \\ e\end{array}\right)_{\!\mathrm{L}}, \qquad 
\mathsf{L}_{\mathrm{R}} = \left(\begin{array}{c} \nu \\ e\end{array}\right)_{\!\mathrm{R}}\;,
\label{eq:lrchirall}
\end{equation}
with quantum numbers $(\mathbf{1},\mathbf{2},\mathbf{1})_{-1}$ and $(\mathbf{1},\mathbf{1},\mathbf{2})_{-1}$.  In the gauge-covariant derivative specifying the interactions, the normalization of the $\mathrm{SU(2)_R}$ gauge coupling $g_{\mathrm{R}}$ is defined in a manner analogous to that for the $\mathrm{SU(2)_L}$ gauge coupling $g_{\mathrm{L}}$. Again we will take the \cgg\ and \wigg\ couplings $g_3$ and $g_{\mathrm{L}} = g$ to retain their standard-model values, and we assume that the $\mathrm{SU(2)_R}$ and \blgg\ gauge couplings are similar in magnitude to $g_{\mathrm{L}}$.

As in the \SMR, at a scale near $\Lambda_{\mathrm{QCD}}$, QCD confines colored objects and spontaneously breaks the global chiral \chiral{2}\ symmetry associated with the massless quarks, leaving a residual $\mathrm{SU(2)_V}$ isospin symmetry. The three would-be pions become the longitudinal components of gauge bosons that correspond to the broken (axial) generators, endowing the massive states with a common mass given by $\overline{M}^2 = (g_{\mathrm{L}}^2 + g_{\mathrm{R}}^2)\bar{f}_\pi^2/4$. If $g_{\mathrm{L}} = g_{\mathrm{R}} = g$, then $\overline{M}$ would be a factor $\sqrt{2}$ larger than the \SMR\ value of  $\overline{M}_{W}$. The corresponding triplet of  $\mathrm{SU(2)_V}$ gauge bosons remains massless, and couples to (vectorial) weak isospin with a gauge coupling
\begin{equation}
g_{\mathrm{V}} = \frac{g_{\mathrm{L}}g_{\mathrm{R}}}{\sqrt{g_{\mathrm{L}}^2 + g_{\mathrm{R}}^2}} .
\label{eq:gvdef}
\end{equation}
If $g_{\mathrm{R}} \approx g_{\mathrm{L}}$, then $g_{\mathrm{V}} \approx g/\sqrt{2} \approx 0.46$. As before, because of the confinement and chiral symmetry breaking, the quarks pick up effective ``constituent'' masses of order $\Lambda_{\mathrm{QCD}}$, and the physical states are color-singlet mesons, baryons, and glueballs.

Below the QCD scale, the gauge symmetry has been broken to
\begin{equation}
\cgg \otimes \mathrm{SU(2)_V} \otimes \blgg ,
\label{eq:lrlesym}
\end{equation}
but since only color-singlet degrees of freedom persist below the confinement scale, it suffices to classify the states according to their $(\mathrm{SU(2)_V})_{B-L}$ quantum numbers. The lepton Dirac field $\mathsf{L} = {\nu \choose e}$ transforms as $(\mathbf{2})_{-1}$, the nucleon doublet $p \choose n$ as $(\mathbf{2})_{1}$, and the V-gluons as $(\mathbf{3})_0$.

The leptons remain massless to this stage, so the effective low-energy theory, with hadrons integrated out, exhibits a global $\mathrm{SU(2)^{(\ell)}_L}\otimes \mathrm{SU(2)^{(\ell)}_R}$ leptonic  chiral symmetry. The $\mathrm{SU(2)_V}$ interaction is asymptotically free,  and hence the associated coupling $g_{\mathrm{V}}$ increases as the energy scale decreases below $\Lambda_{\mathrm{QCD}}$. At a sufficiently low energy scale $\Lambda_{\mathrm{V}}$, where $\alpha_{\mathrm{V}} \equiv g_{\mathrm{V}}^2/4\pi$ approaches unity, the $\mathrm{SU(2)_V}$ interaction confines and breaks the leptonic chiral symmetry. Because the running of $\alpha_{\mathrm{V}}$, calculated by standard renormalization group methods, is logarithmic, the scale $\Lambda_{\mathrm{V}}$ is exponentially smaller than $\Lambda_{\mathrm{QCD}}$, provided that the couplings $g_{\mathrm{L}}(\Lambda_{\mathrm{QCD}})$ and $g_{\mathrm{R}}(\Lambda_{\mathrm{QCD}})$ are small, as we assume. We adopt the QCD (and technicolor) criterion that chiral symmetry breaking occurs in a channel $\mathbf{r} \times \mathbf{\bar{r}} \to \mathbf{1}$ when $\alpha_{\mathrm{V}} C_2(\mathbf{r}) \approx 1$, where $C_2(\mathbf{r})$ is the quadratic Casimir operator , of the fermion representation $\mathbf{r}$  under the gauge group,~\cite{Appelquist:1986an,Appelquist:1988yc}. With the illustrative value $g_{\mathrm{V}}\approx 0.46$ given above, we estimate that $\Lambda_{\mathrm{V}} \approx 10^{-24}\Lambda_{\mathrm{QCD}}$.  Maximally-attractive-channel and vacuum-alignment arguments point to the formation of a lepton condensate,
\begin{equation}
\langle\bar{\mathsf{L}}\mathsf{L}\rangle = \langle (\bar{e}e + \bar{\nu}\nu)\rangle,
\label{eq:lepcon}
\end{equation}
that respects the $\mathrm{SU(2)_V} \otimes \blgg$ gauge symmetry, but breaks the leptonic chiral symmetry to a ``leptonic isospin'' symmetry $\mathrm{SU(2)^{(\ell)}_V}$. In the process, the electron and neutrino acquire equal dynamical masses, of order $\Lambda_{\mathrm{V}}$. At least in principle, therefore, this model avoids the massless deconfined charged fermion pathology of the \SMR.

The process by which the leptons pick up dynamical masses at the scale $\Lambda_{\mathrm{V}}$ is analogous to the one by which the confining \cgg\ interaction produces  dynamical quark masses at the scale $\Lambda_{\mathrm{QCD}}$. The $\mathrm{SU(2)_V}$-singlet bound states are of three types: leptonic mesons, of the form $\ket{(\bar{e}e + \bar{\nu}\nu)}$, with lepton number $L = 0$; leptonic baryons, with $L=2$, and V-glueballs.\footnote{The lowest-lying leptonic mesons are pseudoscalars and vectors; the ground-state leptonic baryons have spin 0 and 1.} The confinement of $\mathrm{SU(2)_V}$ charge means that at distances $r \gtrsim \Lambda_{\mathrm{V}}^{-1}$ hadrons will be (very lightly) bound into $\mathrm{SU(2)_V}$ singlets. On the same scale there would also be $\mathrm{SU(2)_V}$-singlet combinations of nucleons and leptons, including an ersatz $(e~p)$ Hydrogen atom bound by both the $\mathrm{SU(2)_V}$ and \blgg\ gauge interactions. If the second confinement scale $\Lambda_{\mathrm{V}}$ is indeed extremely tiny, then the Bohr radius of a putative atom would be macroscopic, and the integrity of matter would again be lost.

In the limit of low energies $(\lesssim \Lambda_{\mathrm{V}})$, all the physical states are $\mathrm{SU(2)_V}$ singlets, and so we can express the electric charge operator as
\begin{equation}
Q = \cfrac{1}{2}(B-L) .
\label{eq:lrqople}
\end{equation}
In this sense, the surviving Abelian symmetry at low energies is not \emgg, but \blgg. This circumstance is a reminder that the emergence of electromagnetism as a low-energy, long-range interaction is not automatic; it depends not only on the choice of gauge group, but also on the pattern of symmetry breaking.

As a final comment on the \LR\ \textit{Gedanken} world, let us note that three Nambu--Goldstone bosons appear as a consequence of the breaking of the global symmetry 
$\mathrm{SU(2)^{(\ell)}_L}\otimes \mathrm{SU(2)^{(\ell)}_R} \to \mathrm{SU(2)^{(\ell)}_V}$. These massless bosons do not cause any infrared pathology, because they are neutral under the surviving \blgg\ long-range force. Moreover, the NGBs have derivative couplings, so that they become noninteracting in the limit of vanishing energy.

\section{Pati-Salam \textit{Gedanken} World \label{sec:psworld}}
Our focus in this study has been on \textit{Gedanken} worlds in which electroweak symmetry breaking is dominantly caused by the formation of a quark condensate in QCD. Here we take up a generalization of that scenario in the setting of a prototype unified theory of the strong, weak, and electromagnetic interactions, in which the \textit{provocateur} of electroweak symmetry breaking is not \cgg, but a larger gauge group in which QCD is embedded. Specifically, we consider an extension of the  \lrgg\ explored in \S\ref{sec:lrsym}, identifying lepton number as the fourth color, and so combining \cgg\ and \blgg\ into the \psgg\ introduced by Pati \& Salam~\cite{Pati:1974yy}. In the resulting \psew\ \textit{Gedanken} world, it is the unbroken \psgg, rather than \cgg, that produces a fermion condensate that breaks the electroweak gauge symmetry. Like the \LR\ example, this \PS\ model exhibits two widely separated confinement scales and yields no deconfined massless charged fermions.

Considered as a description of the real world, the \psew\ model must be broken to the standard-model gauge group \smgg\ at a high scale of order $1\tev$. That step is required to satisfy the stringent upper bounds on right-handed charged-current interactions, and also to suppress decays such as $K^+ \to \pi^+\mu^+e^-$ that are allowed in the unbroken theory with three fermion generations. The symmetry is conventionally broken by the Higgs mechanism, but it may also be broken dynamically~\cite{Appelquist:2002me}.

The PS model accounts for the quantization of electric charge. The $(B-L)$ operator is proportional to the fourth diagonal generator of the $\mathrm{SU(4)}$ Lie algebra, and so the analogue of \Eqn{eq:qoplrdef} defines the electric charge operator as the sum of three generators, one for each of the non-Abelian groups in \psew.

The chiral quark and lepton fields defined in \Eqn{eq:lrchiralq} and \Eqn{eq:lrchirall} are joined into the fermion representations
\begin{equation}
\mathsf{F}_{\mathrm{L}} = \left( \mathsf{Q}^a, \mathsf{L}\right)_{\!\mathrm{L}}, \qquad \mathsf{F}_{\mathrm{R}} = \left( \mathsf{Q}^a,  \mathsf{R}\right)_{\!\mathrm{L}}\;,
\label{eq:pschiral}
\end{equation}
which transform as the $(\mathbf{4},\mathbf{2},\mathbf{1})$ and $(\mathbf{4},\mathbf{1},\mathbf{2})$ representations of \psew, respectively. 

In light of what has gone before, we can quickly enumerate the major characteristics of the  one-generation \PS\ \textit{Gedanken} world. The \psgg\ interaction is asymptotically free, so as the energy scale decreases through a characteristic scale $\Lambda_\mathrm{PS}$, the gauge coupling becomes strong, the theory confines and produces equal fermion condensates $\langle(\bar{u}u + \bar{\nu}\nu)\rangle$ and $\langle(\bar{d}d + \bar{e}e)\rangle$ that respect the \psgg\ symmetry, but break $\chiral{2} \to \mathrm{SU(2)_V}$. Just as in the \LR\ model, this breaking confers a common mass on the gauge bosons corresponding to the broken axial generators, with the difference that the mass scale is set by the \psgg\ counterpart of $\bar{f}_\pi$. All the fermions gain a common dynamical mass, of order $\Lambda_{\mathrm{PS}}$.

Below the PS scale, the gauge symmetry has been broken to
\begin{equation}
\psgg \otimes \mathrm{SU(2)_V} ,
\label{eq:pslesym}
\end{equation}
but since only \psgg-singlet degrees of freedom persist below the confinement scale, it suffices to classify the states according to their $\mathrm{SU(2)_V}$ quantum numbers. These include the \psgg\ mesons, bosonic baryons, and glueballs. There are no light fermions below the confinement scale $\Lambda_{\mathrm{PS}}$. As the $\mathrm{SU(2)_V}$ interaction becomes strong at very low energies, it produces a spectrum of V-glueballs with masses set by its confinement scale, $\Lambda_{\mathrm{V}}$. Using the coupling constant $g_{\mathrm{V}}$ given in \Eqn{eq:gvdef}, we estimate $\Lambda_{\mathrm{V}} \approx 10^{-21}\Lambda_{\mathrm{PS}}$. In this case, no exact Abelian symmetry remains, so the low-energy effective \PS\ theory has no long-range force. Again a gauge symmetry that might have given rise to low-energy electromagnetism does not, because of the pattern of ``electroweak'' symmetry breaking. 

\section{Conclusion \label{sec:conc}}

In this paper we have studied toy models built upon the \smgg\  standard-model gauge group and fermion content similar to that of the standard model, but without any Higgs field. The electroweak symmetry is hidden, so that $\ewgg \to \emgg$, through the quark condensates that spontaneously break chiral symmetry in quantum chromodynamics with massless quarks.
In these modified standard models, the quarks and leptons do not acquire any masses as a consequence of electroweak symmetry breaking. The agent that endows the fermions with mass need not be the same one responsible for hiding the \ewgg\ symmetry and giving mass to the weak bosons. {Nor is it guaranteed that all fermion masses have a common origin.}

In the absence of (hard) quark masses, the goodness of the vectorial $\mathrm{SU}(n_q)$ flavor symmetry is enhanced, but the broad outlines of the hadron spectrum are left unchanged, with one important exception. In the one-generation modified standard model \SMRn{1}, the pions disappear from the physical spectrum to become the third (longitudinal) components of the electroweak gauge bosons $W^\pm$ and $Z^0$. The vanishing of up- and down-quark masses may leave the proton heavier than the neutron, in which case it would be unstable against $p \to n e^+ \nu_e$ $\beta$ decay, leaving a \textit{Gedanken} world without hydrogen atoms. In any case, the integrity of matter is compromised in \textit{Gedanken} worlds that contain leptons with vanishing or infinitesimal masses.

In the \SMR, the electroweak symmetry breaking scale is set by the QCD
scale, which leads to many interesting and striking differences with respect to the
real world.  Lacking a Higgs boson, the \SMR\ exhibits strong dynamics in $WW$, $WZ$, and $ZZ$ scattering at c.m.\ energies below $1\gev$. Electroweak symmetry breaking on the QCD scale implies a very significant reduction in the weak-boson masses, so that $\overline{M}_W \approx 30\mev$ in the \textit{Gedanken} world. This makes the low-energy weak interactions comparable to the electromagnetic interactions. A corresponding increase in the \SMR\ value of the Fermi constant dramatically accelerates $\beta$ decay and enhances neutrino-nucleon cross sections. Moreover, weak interactions and residual strong interactions are of comparable strength, so the systematics of nuclear forces are changed in important ways. In particular, nuclear forces of weak origin would be \textsf{P}- and \textsf{C}-violating. On the other hand, vanishing quark masses would obviate the strong \textsf{CP} problem.

Models with vanishing or infinitesimal masses for unconfined charged leptons are pathological, being unstable against the breakdown of the vacuum and the creation of a plasma of electron-positron pairs. It is therefore appealing to consider the modified standard model as a low-energy effective field theory, and to investigate ultraviolet completions that induce fermion mass terms that are hard on the QCD scale.  

In
particular, we contrasted the properties of two types of models, one that has,
in isolation, no fermion bare mass terms and pathological infrared behavior
due to the unconfined massless leptons, and another that has bare
fermion mass terms, with magnitudes bounded by the requirement that the model
can serve as a reasonable effective theory up to energies somewhat above the
hadronic scale.  

The generalization to $N_{\mathrm{c}} \ne 3$ exposes another connection between quantum chromodynamics and the electroweak sector. In the standard model, with the Fermi constant a fixed parameter, the weak-boson masses scale as $1/\sqrt{N_{\mathrm{c}}}$, whereas in the \SMR, weak-boson masses are independent of the number of colors, but $\overline{G}_{\mathrm{F}} \propto 1/{N_{\mathrm{c}}}$.

Two no-Higgs modifications of extensions to the standard-model gauge group display interesting characteristics. In the first of these, we considered a left-right--symmetric theory based on \lrgg\ gauge symmetry. The spontaneous breaking of the chiral symmetry associated with massless quarks that occurs near the color-confinement scale $\Lambda_{\mathrm{QCD}}$ breaks the gauged \chiral{2}\ to a residual exact $\mathrm{SU(2)_V}$ gauge symmetry, and gives masses to the \chiral{2}\ weak bosons. The $\mathrm{SU(2)_V}$ gauge symmetry is asymptotically free, and so confines leptons at a characteristic scale, $\Lambda_{\mathrm{V}} \ll \Lambda_{\mathrm{QCD}}$. The two scales of confinement and chiral symmetry breaking are widely separated provided that the weak gauge couplings are small enough that \chiral{2}$\otimes \mathrm{U(1)}_{B-L}$ may be considered as a perturbation on the \cgg\ strong interaction. Because leptons are confined, this model avoids the problem of massless deconfined charged particles. A long-range interaction coupled to $B-L$, rather than electric charge, remains after the confining interactions have accomplished their symmetry breaking. Electromagnetism is thus not an automatic outcome.

Our final example was a prototype unified theory that incorporates \psgg\ gauge symmetry, in which \cgg\ and $\mathrm{U(1)}_{B-L}$ are joined by regarding lepton number as the fourth color. One of the virtues of the resulting \psew\ gauge symmetry is that it provides a natural explanation of the quantization of electric charge. At a characteristic scale $\Lambda_{\mathrm{PS}}$, the \psgg\ interaction confines and induces the breaking of the gauged \chiral{2}\ to a residual exact $\mathrm{SU(2)_V}$ gauge symmetry. Since both quarks and leptons are confined, this model avoids the pathology of massless charged deconfined fermions. The residual $\mathrm{SU(2)_V}$ is asymptotically free, so at a (very low) characteristic scale $\Lambda_{\mathrm{V}}$ it confines the gauge bosons of this interaction, V-gluons, into $\mathrm{SU(2)_V}$-singlet V-glueballs. In contrast to the no-Higgs versions of the standard model and the left-right symmetric theory, no $\mathrm{U(1)}$ gauge symmetry survives, and so---leaving aside gravity---there is no long-range force at all. 

One of the goals of this study was to explore how different the physical world would be, if there were nothing like the Higgs mechanism to hide the electroweak symmetry. We regard that as important preparation for interpreting the insights we hope to derive from experiments at the Large Hadron Collider. The \SMR\ provides an explicit theoretical laboratory in which to study how properties of our world would change if the scale of electroweak symmetry breaking varied from its real-world value. The approach taken here complements studies that retained a Higgs mechanism but allowed variations in the weak scale $v$. 

\textit{Gedanken} worlds also help us to take a fresh look at the real world and the way we have understood it. Since we have not definitively established the source of electroweak symmetry breaking, it is useful to ask how it might deviate from the standard hypothesis. The early technicolor models were based on a close analogy with QCD, and the general idea of dynamical electroweak symmetry breaking merits continued study. Even when we do include a Higgs mechanism that gives rise to a
   realistic scale for electroweak symmetry breaking, it is important to bear in mind that there are at least two
   contributing sources---even if the QCD source is but a tiny effect.  This serves as a reminder  that there can be several contributions,
   at quite different mass scales, to the breaking of a given symmetry.

\begin{acknowledgments}

We thank Peter Arnold, James Bjorken, Stan Brodsky, Robert Cahn, Gerard 't Hooft, Andreas Kronfeld, Gerald Miller, Shmuel Nussinov, Mary Hall Reno, Bert Schellekens, and Nikolai Uraltsev for stimulating comments, discussions, questions, and advice.  Fermilab is operated by the Fermi Research Alliance under contract no.\  DE-AC02-07CH11359 with the
U.S.\ Department of Energy.  The research of RS was partly supported by
NSF-PHY-06-53342.  Part of this research was performed during his visit  to
the Fermilab theory group. He thanks the theory group for hospitality and
the Universities Research Association for support as a URA Visiting Scholar. CQ thanks Hans K\"{u}hn and Uli Nierste for a stimulating environment in Karlsruhe and acknowledges with pleasure the generous support of the Alexander von Humboldt Foundation. He is grateful to Luis \'{A}lvarez-Gaum\'{e} and other members of the CERN Theory Group for their hospitality.

\end{acknowledgments}

\appendix
\section{Neutron-proton Mass Difference  \label{app:npdiff}}
The problem of calculating hadron masses has received much attention since the rich spectrum of strongly interacting particles began to reveal itself in the 1960s. The state of the art has progressed from early considerations in the framework of the nonrelativistic quark model~\cite{Kokkedee} through relativistic formulations such as the MIT bag model~\cite{Chodos:1974je} and the Bethe-Salpeter equation~\cite{Salpeter:1951sz}, to methods based on light-front wave functions and the Anti-de Sitter/conformal field theory connection~\cite{deTeramond:2005su} and lattice gauge theory computations~\cite{Lattice2008}. With respect to so-called electromagnetic mass differences---splittings within an isospin multiplet---there has been parallel progress from the Cottingham formula~\cite{Cottingham} to \textit{ab initio} calculations in lattice QCD~\cite{Duncan:1996be,Beane:2006fk}. 

Our modest goal in this Appendix is to expose two differences between the real world and the \SMR\ \textit{Gedanken} world: the absence in the \SMR\ of quark masses, and the appearance in the \SMR\ of weak contributions to the neutron-proton mass difference that are negligible in the real world. In the real world, the difference between up- and down-quark masses increases the neutron mass relative to the proton mass. The weak-interaction contributions have a similar effect in the \SMR.

It will be sufficient for our purposes to examine the sources of the $n$-$p$ mass difference within the quark model, using as a starting point the $\mathrm{SU(6)}$ flavor-spin wave functions,
\begin{eqnarray}
\ket{p\uparrow} & = & (1/\sqrt{18})\left( 2\uup\ddn\uup - \udn\dup\uup - \uup\dup\udn \right. \nonumber \\
 & & -\dup\udn\uup + 2\ddn\uup\uup - \dup \uup\udn \nonumber \\
  & & \left. - \uup\udn\dup - \udn\uup\dup + 2\uup\uup\ddn\right) , \nonumber \\[-6pt]
  & & \\[-6pt]
  \ket{n\uparrow} & = & -(1/\sqrt{18})\left(2\dup\udn\dup - \ddn\uup\dup - \dup\uup\ddn \right. \nonumber \\
   & & -\uup\ddn\dup + 2 \udn\dup\dup - \uup\dup\ddn \nonumber \\
   & & \left. - \dup\ddn\uup - \ddn\dup\uup + 2\dup\dup\udn\right) , \nonumber
\end{eqnarray}
here written for proton and neutron with spin up.

Within a multiplet, a standard analysis identifies three distinct contributions to mass differences: (i) a difference between the masses of the up and down quarks, (ii) pairwise Coulomb interactions among the valence quarks, and (iii) pairwise hyperfine or magnetic-moment interactions among quarks. A plausible expression for particle masses within a real-world multiplet is then (\cf~\cite{Quigg:1981nb})
\begin{eqnarray}
M & = & M_0 + n_d(m_d + \delta_d) + n_u(m_u + \delta_u) +\! \left\langle\frac{\alpha}{r}\right\rangle\!\!\sum_{i<j}Q_iQ_j \nonumber \\
 & &\qquad -  \frac{8\pi}{3}\abs{\Psi_{ij}(0)}^2\!\!\left\langle \sum_{i<j}\mu_i\mu_j\vec{\sigma}_i \cdot \vec{\sigma}_j \right\rangle\;. 
\label{eq:multimass}
\end{eqnarray}
In this equation, $n_i$ is the number of quarks of flavor $i$ in the hadron, $\delta_i$ is the electromagnetic correction to the constituent-quark mass of flavor $i$, $Q_i$ and $\mu_i$ are the electric charge and magnetic moment of quark $i$, and $\Psi_{ij}(0)$ is the wave function at zero separation of quarks $i$ and $j$. The symmetry of the wave function has been exploited in writing the Coulomb term. 

The formula \Eqn{eq:multimass} incorporates different aspects of the dynamics contributing to the nucleon mass and entails a number of approximations. Fits to the masses and magnetic moments of the baryon octet require that the quark magnetic moments be inversely proportional to the constituent-quark, not current-quark) masses~\cite{De Rujula:1975ge}. Questions of internal consistency arise because the chromomagnetic hyperfine term is a short-distance phenomenon, and at short distances the quarks behave as quasifree particles with no dynamically generated masses~\cite{Nussinov:2008rm}. With such caveats in mind, we proceed with the analysis.

Without appealing to a specific dynamical model to compute the mean quark separation $\langle r^{-1} \rangle$ and $\abs{\Psi_{ij}(0)}^2$, we may recast \Eqn{eq:multimass} in terms of Coulomb and magnetic contributions $\delta M_{\mathrm{C}}$ and $\delta M_{\mathrm{M}}$ as \begin{eqnarray}
M & = & M_0 + n_d(m_d + \cfrac{1}{9}\delta) + n_u(m_u + \cfrac{4}{9}\delta)\label{eq:multimassp} \\
 & & +\delta M_{\mathrm{C}} \sum_{i<j}Q_iQ_j  + \delta M_{\mathrm{M}} \left\langle \sum_{i<j}Q_iQ_j\vec{\sigma}_i \cdot \vec{\sigma}_j \right\rangle\;, \nonumber
\end{eqnarray}
where we have assumed that the magnetic moments are proportional to the quark charges and the corrections to the constituent-quark masses are proportional to the quark charges squared. The electromagnetic mass shift $\delta = \alpha M_0/3\pi$, times a constant of order unity, so approximately $1\mev$, comparable in magnitude to the terms with which it is grouped.

Evaluating \Eqn{eq:multimassp}, we find
\begin{eqnarray}
M_p & = & M_0 + (m_d+2m_u) + \;\,\,\delta\qquad\qquad +\cfrac{4}{3}\delta M_{\mathrm{M}} \nonumber \\
 & &   \label{eq:pandn}\\
 M_n & = & M_0 + (2m_d+m_u)+ \cfrac{2}{3}\delta -\cfrac{1}{3}\delta M_{\mathrm{C}} + \delta M_{\mathrm{M}}\;, \nonumber
 \end{eqnarray}
 so that, in the real world,
 \begin{equation}
 M_n - M_p = (m_d-m_u)  -\cfrac{1}{3}(\delta + \delta M_{\mathrm{C}} + \delta M_{\mathrm{M}}).
 \label{eq:npdiff}
 \end{equation}
 The current-quark mass difference, $(m_d-m_u) > 0$, outweighs the negative contribution of the other three terms, leading to the observed $M_n - M_p \approx 1.293\mev$~\cite{Amsler:2008zz}. If we were to set $m_u$ and $m_d$ to zero, but otherwise retain all the features of the real world, the proton would outweigh the neutron.
 
Simply erasing the current-quark masses from  \Eqn{eq:npdiff} is not sufficient to describe the $n$-$p$ mass difference in the \SMR, in which the strength of weak interactions is greatly enhanced over the real-world level. In the \textit{Gedanken} world, we must take into account contributions due to the exchange of virtual $W$ and $Z$ bosons. In the spirit of the quark model, these arise from the difference between constituent-quark mass shifts due to $Z$ emission and reabsorption for the $d$ and $u$ quarks and the difference between $Z$-exchange diagrams for $dd$ scattering in the neutron and $uu$ scattering in the proton, all other $W$- and $Z$-exchange contributions being in common. The two contributions (viz., $Z$ exchange between different quark lines and corrections to individual quark propagators due to $Z$ emission and reabsorption) share some similarities. Accordingly, we can anticipate the character of a detailed evaluation by estimating the $Z$-exchange contribution in the static limit of zero momentum transfer as
\begin{equation}
\left.\overline{M}_n - \overline{M}_p\right|_{Z\mathrm{-ex}} = \frac{\overline{G}_{\mathrm{F}}\Lambda_h^3}{2\sqrt{2}}\left[(L_d^2 + R_d^2) - (L_u^2 + R_u^2)\right]
\label{eq:weaknp}
\end{equation}
where $\Lambda_h$ is a typical hadronic scale and the chiral couplings of $Z$ to quark $q$ are
\begin{equation}
L_q = \tau_3 - 2Q_q x_{\mathrm{W}},\; R_q = - 2Q_q x_{\mathrm{W}}
\end{equation}
with $\tau_3$ twice the weak isospin projection of $q$ and $x_{\mathrm{W}} \equiv \sin^2\theta_{\mathrm{W}}$ the weak mixing parameter. {Note that in the real world, this contribution is indeed entirely negligible: with the plausible value $\Lambda_h = f_\pi \approx 100\mev$, $G_{\mathrm{F}}\Lambda_h^3 \approx 10^{-5}\mev$.} Inserting the \SMRn{1} \textit{Gedanken}-world value $\overline{G}_{\mathrm{F}}/\sqrt{2} = 1/2\bar{f}_\pi^2$, we compute
\begin{equation}
\left.\overline{M}_n - \overline{M}_p\right|_{Z\mathrm{-ex}} = \frac{\Lambda_h^3}{3\bar{f}_\pi^2} x_{\mathrm{W}}(1 - 2x_{\mathrm{W}}) \approx  \frac{\Lambda_h^3}{24\bar{f}_\pi^2}.
\label{eq:npweak}
\end{equation}
where we approximated $x_{\mathrm{W}} = \cfrac{1}{4}$, close to the real-world value, in the last step. This contribution is positive and plausibly measured in MeV, so could well tip the balance back in favor of the neutron as the heavier nucleon. {That electromagnetic and weak contibutions have opposite sign, so long as $x_{\mathrm{W}} < \cfrac{1}{2}$, was noted long ago in the context of the Cottingham formula for real-world mass differences~\cite{Love:1973ab}.}


\end{document}